\documentclass[twocolumn,twocolappendix]{aastex6}
\usepackage{amssymb,amsmath}	
\usepackage{graphicx}
\usepackage{xcolor}
\usepackage{hyperref}

\newcommand\abs[1]{\left|#1\right|}

\newcommand{\Mo}{\rm{M}_\odot}

\mathchardef\mhyphen="2D
\newcommand{\corr}{\mathrm{Corr}}
\renewenvironment{description}[1][0pt]
  {\list{}{\labelwidth=0pt \leftmargin=#1
   }}
  {\endlist}
  
  \defcitealias{2016ApJ...830...31B}{Bellinger \& Angelou et al. \textcolor{black}{(}2016\textcolor{black}{, BA1 hereinafter)} }

\shorttitle{Statistical Properties of the MS}
\shortauthors{Angelou \& Bellinger et al.}

\begin{document}


\title{On The Statistical Properties of the Lower Main Sequence}


\author{George C. Angelou \altaffilmark{1,2}, Earl P. Bellinger\altaffilmark{1,2,3,4}, Saskia Hekker\altaffilmark{1,2} and Sarbani Basu\altaffilmark{4}}
\affil{\altaffilmark{1} Max-Planck-Institut f\"{u}r Sonnensystemforschung, Justus-von-Liebig-Weg 3, 37077 G\"{o}ttingen, Germany\\
\altaffilmark{2} Stellar Astrophysics Centre, Department of Physics and Astronomy, Aarhus University, Ny Munkegade 120, DK-8000 Aarhus C, Denmark \\
\altaffilmark{3} Institut f\"{u}r Informatik, Georg-August-Universit\"{a}t G\"{o}ttingen, Goldschmidtstrasse 7, 37077 G\"{o}ttingen, Germany \\
\altaffilmark{4} Department of Astronomy, Yale University, New Haven, CT 06520, USA}


\begin{abstract}
Astronomy is in an era where all-sky surveys are mapping the Galaxy. 
The plethora of photometric, spectroscopic, asteroseismic and astrometric data allows us to characterise the comprising stars in detail.  
Here we quantify to what extent precise stellar observations reveal information about the properties of a star, including properties that are unobserved, or even unobservable. 
We analyse the diagnostic potential of classical and asteroseismic observations for inferring stellar parameters such as age, mass and radius from evolutionary tracks of solar-like oscillators on the lower main sequence. 
We perform rank correlation tests in order to determine the capacity of each observable quantity to probe structural components of stars and infer their evolutionary histories. We also analyse the principal components of classic and asteroseismic observables to highlight the degree of redundancy present in the measured quantities and demonstrate the extent to which information of the model parameters can be extracted.
We perform multiple regression using combinations of observable quantities in a grid of evolutionary simulations and appraise the predictive utility of each combination in determining the properties of stars.
We identify the combinations that are useful and provide limits to where each type of observable quantity can reveal information about a star. We investigate the accuracy with which targets in the upcoming \emph{TESS} and \emph{PLATO} missions can be characterized.  We demonstrate that the combination of observations from \emph{GAIA} and \emph{PLATO} will allow us to tightly constrain stellar masses, ages and radii with machine learning for the purposes of Galactic and planetary studies.
\end{abstract}



\keywords{methods: statistical --- stars: abundances --- stars: fundamental parameters --- stars: low-mass --- stars: oscillations --- stars: solar-type}


\section{Introduction} 

The main sequence is generally considered the most well-understood phase of stellar evolution. 
Our Sun is a main-sequence star, and its proximity provides a wealth of constraints to the physics that may occur in low-mass counterparts during this phase \citep{2015SSRv..196...49B,2016LRSP...13....2B}. 
Core-hydrogen burning stars are long-lived and hence numerous: indeed, the majority of the stars for which we can resolve parallaxes reside on the main sequence \citep{2016arXiv160904172G}. 
Additionally, many stars of this type display stochastic or ``solar-like'' oscillations that serve to reveal the stellar interior (see for example \citealt{2013ARA&A..51..353C} for a review on solar-like oscillators).  
Main-sequence stars are important astrophysical laboratories for testing theories of stellar physics, structure, and evolution; and are a testbed for general physical theories such as nuclear fusion, diffusion, and convection \citep{1994MNRAS.269.1137B,1990ARAA..28..263S}. 

Despite all of this, however, the ages of main-sequence stars remain uncertain to at least 10\%. This uncertainty stems not only from observational imprecision, but also from the inability of observations to fully constrain stellar parameters.  
Recently, \citetalias{2016ApJ...830...31B} showed that even for stellar models without observational uncertainties, some model attributes of stars---such as their initial helium abundance or efficiency of convection---could not be fully resolved via global information that can be gleaned from their surfaces.

It is well-known that different observable quantities of stars constrain different model properties. For example, in the now-famous Christensen-Dalsgaard diagram (C-D diagram, the so-called ``asteroseismic H-R diagram''), in which the large frequency separation is plotted against the small frequency separation (Appendix \ref{sec:sdefs}), the large frequency separation covaries with the mass of the star and the small frequency separation covaries with its core-hydrogen abundance. Hence, observing one of these quantities sheds light on its unobservable counterpart.
However, to date, a systematic investigation of the extent to which each observable quantity constrains each model property has not been performed.

The equations dictating stellar structure and evolution, and the corresponding microphysics that these equations respond to, give rise to emergent behaviors that are difficult to characterize through examination of the constituting ingredients themselves. To elucidate these opaque relationships, we seek to determine the extent to which observable stellar properties are capable of constraining the internal structures, chemical mixtures, and evolutionary histories of stars. Here we employ the methodology of exploratory data science, a statistical philosophy by which underlying structure in data --- simulated or otherwise --- is unearthed.

BA1 used machine learning to build a statistical description of main-sequence stellar evolution. They trained a random forest (RF) of decision trees to learn the relationships that exist between model input parameters and their resultant observable quantities. The technique was developed with particular focus on the determination of stellar ages.  Ages are essential for understanding stellar evolution, characterising extrasolar planetary systems and advancing models of galactic chemical evolution. 
Notably, the RF developed by BA1 was able to accurately predict stellar properties such as radii and luminosities using other information collected from the stars in their sample. This illustrates that there is redundant information in the stellar quantities, and that there exist model covariances between these quantities that can be characterized and exploited.

The philosophy employed in BA1 is a departure from the standard practice of stellar model fitting. Ordinarily, stellar parameters of observed stars are sought via $\chi^2$-minimization.
The difference in approaches give rise to two points that motivate this paper:
\begin{enumerate}

    \item Methods based on $\chi^2$-minimization assume that each bit of observed information contributes to the objective of constraining the model properties of a star in an exact proportion to how precisely it has been measured. However, two quantities may be measured independently with no measured covariance, and yet still provide redundant information about the star. The result of such a minimization procedure will therefore be a model that is biased towards that redundant information.  The RF developed in BA1, on the other hand, uses the process of statistical bagging to avoid over-fitting the data (see also \citealt{hastie2005elements}). Here we demonstrate the degree to which the observables  carry redundant information about the star. 

    \item The 
    optimization searches of iterative model finding procedures provide solutions but do not indicate the elements that were important in doing so. The use of regression requires that the observables correlate with those model parameters that we wish to infer.  We therefore identify to what extent each observable constrains each model property, and how well the observables must be measured to achieve a desired precision from the regression.  
    
\end{enumerate}

The method developed in BA1 makes use of an artificial intelligence strategy known as supervised learning. The RF that they train seeks relations in evolutionary simulations that enable model properties to be inferred as precisely as possible. Although the RF performs the analysis quickly, precisely, and automatically; supervised machine learning strategies do not provide much insight into how the end result is obtained. The algorithm essentially produces a formula for inferring stellar properties from observations, but one that is too complex for people to use analytically by hand. 

Here we incorporate a complementary strategy. We use the counterpart of supervised learning---\emph{unsupervised learning}---to explicitly uncover the relations between observable properties of stars and their model parameters. Hence, BA1 is of a strictly practical nature: stellar parameters can be inferred rapidly without regard for the how or why; and this paper is aimed to further an understanding of the processes actually involved in such a deduction. 

In this study we draw heavily from the work presented in BA1.
Our analysis initially focuses on elucidating the inherent statistical properties of the grid of stellar models used to train the BA1 RF. 
We determine the relationships and covariances between a chosen subset of stellar parameters and asteroseismic quantities (see Table \ref{tab:parmdefs}). 
We carry out simultaneous rank correlation tests on the chosen parameters and identify the necessary, dispensable, and irrelevant information for determining each stellar property.
Then, using principle component analysis 
we reduce the dimensionality of the observable quantities and identify to what extent they reveal information of the model parameters.
We subsequently shift the focus of our analysis to how the grid properties are used by the RF and how the choices in the parameters impact on the precision of the regression. 
We train RFs using all combinations of observable quantities in our dataset. The purpose of this is two-fold: first, it is often the case that we wish to quickly characterise a star from a few easily observed quantities---the Hertzsprung-Russell (HR) diagram serves as the classic example. Training and scoring all possible RF combinations provides a means to \emph{quantify} the utility and predictive power of classical and asteroseismic parameters for inferring stellar properties. Secondly, it provides insight into the relationships determined by machine learning algorithms. 
Finally, we identify the observational accuracy required to satisfactorily constrain key stellar parameters. We investigate the observable quantities independently as well as consider the measurements expected from the upcoming \emph{TESS} and \emph{PLATO} missions.

\section{Stellar Models and Parameters}
\begin{table}
{
\renewcommand{\arraystretch}{1.2}
\centering
\begin{tabular}{lll} 
\hline \hline  
\textbf{Qty} & \textbf{Definition} & \textbf{Unit} \\ 
\hline
\multicolumn{3}{l}{Model Input Parameters} \\ 
$M$ & Initial mass & M$_{\odot}$ \\
Y$_0$ & Initial helium mass fraction & \\
Z$_0$ & Initial metal mass fraction & \\
$\alpha_{\rm{MLT}}$ & Mixing length parameter & \\
$\alpha_{\rm{ov}}$ & Overshoot parameter & \\ 
$D$ & Diffusion efficiency factor & \\[8pt] 
\multicolumn{3}{l}{Stellar Attributes} \\
$\tau$ & Age & yr \\
$\tau_{\rm{MS}}$ & Normalised main-sequence lifetime &  \\
M$_{\rm{cc}}$ & Convective core mass & M$_{\odot}$ \\
X$_{\rm{surf}}$ & Surface hydrogren mass fraction & \\
Y$_{\rm{surf}}$ & Surface helium mass fraction & \\
X$_c$ & Central hydrogen mass fraction & \\
$L$ & Luminosity & L$_{\odot}$ \\ 
$R$ & Radius & R$_{\odot}$ \\[8pt]
\multicolumn{3}{l}{Classical Observables} \\
$[\rm{Fe/H}]$ & Surface metallicity & \\
$\log \rm{g}$ & Logarithmic surface gravity &  \\ 
$T_{\rm{eff}}$ & Effective temperature & K \\[8pt]

\multicolumn{3}{l}{Asteroseismic Observables} \\
$\nu_{\rm{max}}$ & Frequency of maximum oscillation power & $\mu$Hz \\
$\langle\Delta\nu_0\rangle$ & Large frequency separation ($\ell=0$) & $\mu$Hz \\
$\langle\delta\nu_{02}\rangle$ & Small frequency separation ($\ell=0,2$) & $\mu$Hz \\
$\langle\delta\nu_{13}\rangle$ & Small frequency separation ($\ell=1,3$) & $\mu$Hz \\
$\langle r_{02}\rangle$ & Frequency separation ratio ($\ell=0,2$) &  \\
$\langle r_{13}\rangle$ & Frequency separation ratio ($\ell=1,3$) &  \\
$\langle r_{01}\rangle$ & Frequency average ratio ($\ell=0,1$) & \\
$\langle r_{10}\rangle$ & Frequency average ratio ($\ell=1,0$) & \\[8pt]\hline 
\end{tabular}
}
\caption{Definitions of the quantities analyzed in this study separated into four parts: model input parameters, stellar attributes, classical observables, and asteroseismic observables. 
Asteroseismic definitions are in Appendix \ref{sec:sdefs}. 
Angled parenthesis indicate the quantity is a calculated weighted median.} 
\label{tab:parmdefs} 
\end{table}

We used  \emph{Modules for Experiments in Stellar Astrophysics} \citep[MESA,][]{Paxton2011} to generate a grid of stellar evolutionary sequences initially for the purpose of training a random forest. The tracks are varied in initial mass $M$, helium Y$_0$, metallicity Z$_0$, mixing length parameter $\alpha_{\text{MLT}}$, overshoot coefficient $\alpha_{\text{ov}}$, and atomic diffusion multiplication factor $D$ (see BA1 \textsection 2.1 for details).
Initial model parameters were chosen in a quasi-random fashion from the parameter ranges listed in Table \ref{tab:prange}.
In total 5325 evolutionary tracks were evolved from ZAMS to either an age of $\tau=15$ Gyr or until terminal-age main sequence (TAMS), which we define as having a fractional core-hydrogen abundance $X_{\text{c}}$ below $10^{-3}$. We conduct our analysis on a subset of stellar models chosen from each sequence so not to bias our statistics towards longer lived stars or numerically challenging evolutionary tracks. 
Details of the choice of input physics, grid generation strategy, and model selection procedure are further outlined in BA1. 
In addition to computing the stellar structure we post process  each model with  the ADIPLS pulsation package \citep{2008Ap&SS.316..113C}.  P-mode oscillations up to spherical degree $\ell=3$ below the acoustic cut-off frequency are computed, and from these, frequency separations and separation ratios calculated (see Appendix \ref{sec:sdefs} for mathematical definitions).

There are many quantities that could be included in the current analysis. 
The 25 parameters we have selected to investigate are listed in Table \ref{tab:parmdefs}. 
They  comprise key asteroseismic and structural quantities and reflect our focus on characterising the relationships between observable quantities (observables hereinafter) and those variables that allow us to generate detailed stellar models.

\begin{table}[ht]
\begin{tabular}{llll}
\hline \hline
Parameter	&	Min Value	&	Max Value	&	Variation 	\\ 
\hline 
Mass	&	0.7	&	1.6	&	linear\\
Y$_0$	&	0.22	&	0.34	&	linear\\
Z$_0$	&	$10^{-5}$	&	$10^{-1}$	&	logarithmic\\
$\alpha_{\rm{MLT}}$	&	1.5	&	2.5	&	linear\\
$\alpha_{\rm{ov}}$	&	$10^{-4}$	&	1	&	logarithmic\\
D	&	$10^{-6}$	&	$10^{2}$	&	logarithmic	\\
\hline
\end{tabular}
\caption{Ranges and sampling strategy for the initial model parameters in the BA1 grid.}
\label{tab:prange}
\end{table}

We consider two parameters not included in the RF training data. 
BA1 elected to omit the frequency of maximum oscillation power, $\nu_{\rm{max}}$ (Equation \ref{equ:nmax}), in their regression model\footnote{$\nu_{\rm{max}}$ does have some role in the algorithm developed by BA1, as it is responsible for the location of the Gaussian envelope used to weight and derive averaged/median frequency separations.}. This quantity displays a strong correlation with $\langle\Delta\nu_0\rangle$  (see Figure \ref{fig:filt_corr}  or \citealt{2009A&A...506..465H,2009MNRAS.400L..80S}) and thus offers very little additional information when frequencies are known. 
We include it in the current analysis because $\nu_{\rm{max}}$ is the simplest global asteroseismic parameter to extract from time-series observations, and because recent work by Theme$\ss$l et al. (private communication) indicates that the $\nu_{\rm{max}}$ scaling relation more accurately reproduces stellar parameters in well-constrained binary systems than the $\langle\Delta\nu_0\rangle$ relation (Equation \ref{equ:dnu}). This is despite the fact that $\langle\Delta\nu_0\rangle$ can be measured more precisely and that the relation can be corrected for temperature and metallicity dependencies (Equation \ref{eq:corrfunc2}) to yield greater accuracy \citep{2016MNRAS.460.4277G, 2016ApJ...822...15S}. 

To complement $\tau$, we have also added normalised main-sequence age, $\tau_{\rm{MS}}$, which describes how parameters change as a function of stellar evolution. 
Many low-mass stars in the grid do not reach the terminal-age main sequence (TAMS) before their evolution is stopped. 
Their main-sequence lifetime is estimated by linearly extrapolating the rate at which the central hydrogen is depleted,
\begin{equation}
\tau_{\rm{TAMS}} =  \frac{\tau_{\rm{last}}}{1-(X_{\rm{c, last}}/X_{\rm{c, init}})}
\end{equation}
where $\tau_{\rm{TAMS}}$ is the TAMS age,  $\tau_{\rm{last}}$ is the age of the last model in  the track,  $X_{\rm{c, last}}$ is the corresponding core-hydrogen abundance for that model and $X_{\rm{c, init}}$ is the core-hydrogen abundance of the initial model in that track. 
For the longest-lived stars we find such an extrapolation is within about 25\% of the true TAMS age. The uncertainty in the extrapolation for these stars stems from the fact we only capture the hydrogen depletion in the early part of the main sequence i.e., when  $X_{\rm{c, last}} > 0.3$.  Estimating the TAMS age in this manner, however,  will not impact our conclusions.
Large discrepancies are limited to a small number of tracks (192) and differences between the true and extrapolated ages are reduced as $X_{\rm{c, last}} \to 0$. 
Main sequence lifetime  provides insight into the general correlations that develop as a function of main-sequence stellar evolution. Thus it is the monotonicity of $\tau_{\rm{MS}}$ within a given track that is key.  The stellar age parameter, on the other hand, is useful for exploring correlations across the whole parameter space.

\section{Rank Correlation Test}
\label{sec:RCT}

\begin{figure}
    \centering
    \includegraphics[width=\columnwidth]{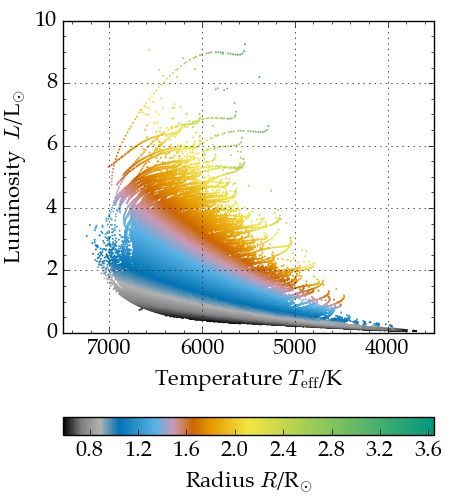}
    \caption{Hertzsprung-Russell diagram for those tracks in the truncated grid (see text for details). Here each model is coloured by stellar radius.}
    \label{fig:HRDRad}
\end{figure}

\begin{figure*}
    \centering
    \includegraphics[trim={1.5cm 0 2cm 1cm},clip, width=\textwidth]{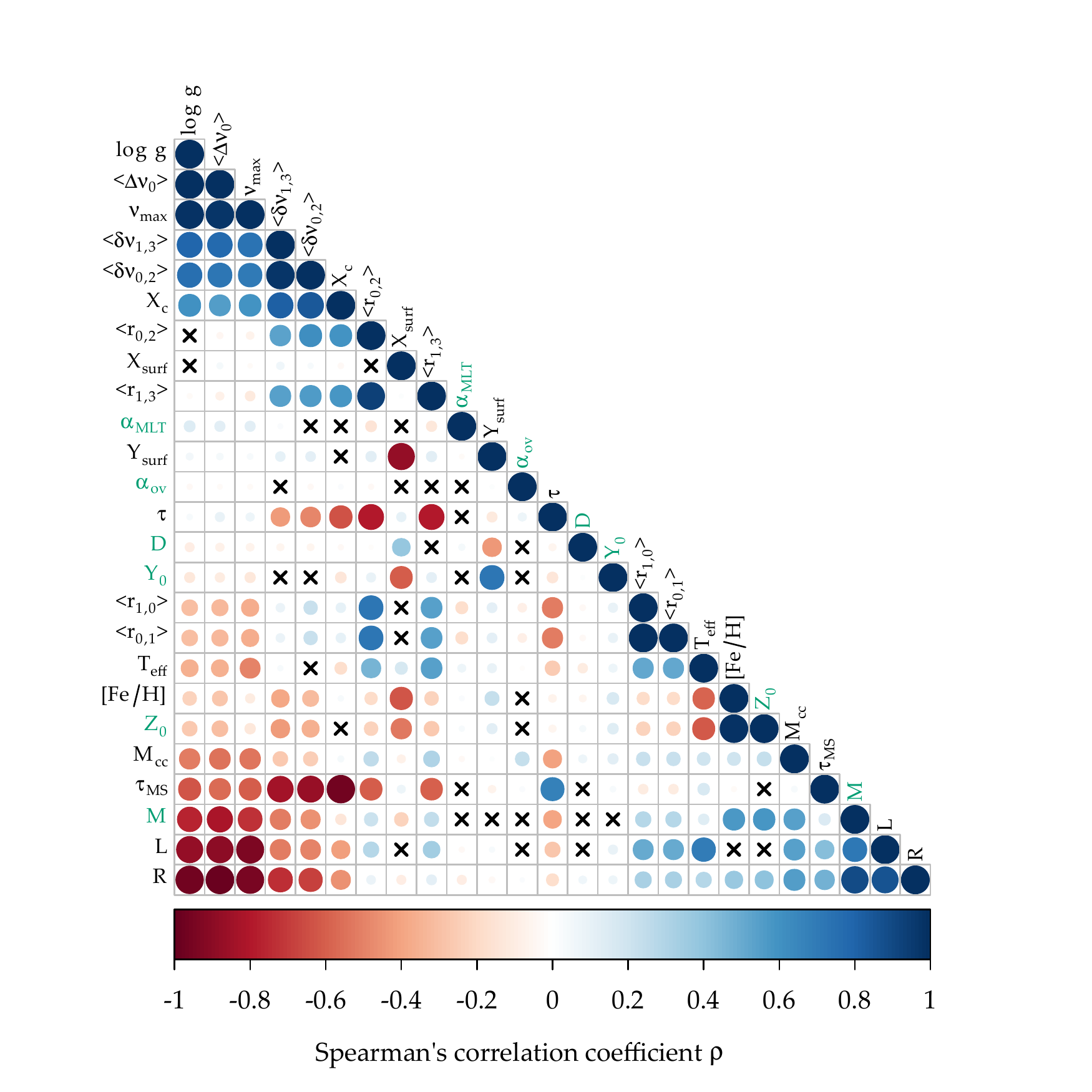}
    \caption{Spearman rank correlation matrix comprising various stellar and asteroseismic parameters. The quantities are as described in Table \ref{tab:parmdefs}
    with model input parameters marked in green.
    The size and the color of each circle both indicate the magnitude of the Spearman coefficient with red and blue denoting negative and positive correlations respectively.  The presence of a cross indicates that 
the two parameters have failed our significance test; i.e., the correlation is indistinguishable from nil. 
The variables are ordered according to their correlation with the first eigensolution of the correlation matrix$^2$.}
    \label{fig:filt_corr}
\end{figure*}

\footnotetext{As principle component analysis is the eigensolution of the correlation (or covariance) matrix, the first eigenvalue indicates the maximum variance in the variables that can be accounted for by a linear model with a single underlying `factor'.
Ordering the parameters in this way  demonstrates the direction of the first principle component (PC$_1$) vector. Figure \ref{fig:filt_corr} thus  offers a visual representation of principle component analysis which we employ in \S \ref{sec:PCA}.}

We begin our analysis with a rank correlation test, the purpose of which being to understand the 
statistical properties of the collective lower main sequence.
This is distinct from typical analyses that focus on the evolutionary properties within 
individual stellar tracks or chemically homogeneous isochrones. By identifying correlations present across the entire parameter space we reveal exploitable relationships available to model fitting and regression methods.

Since many quantities (see Table \ref{tab:parmdefs}) are known to vary in a highly non-linear fashion, we opt to study \emph{rank} statistics. In particular, we replace each quantity by its rank, i.e., an integer representing how big or small a particular quantity is compared to the other models; and calculate Spearman's correlation coefficient $\rho$ between all variables. We further calculate the significance of these correlations (p-values) using the Spearman $\rho$ test. We adopt a conservative significance cut-off of $\alpha = 10^{-5}$ and use the Bonferroni correction to account for the fact that we are making multiple comparisons (625 comparisons). 

This analysis allows us to determine whether quantities vary monotonically in the same direction ($\rho \approx 1$), i.e.~both increasing or both decreasing; monotonically apart ($\rho \approx -1$), i.e.~one increases while the other decreases; or neither ($\rho \approx 0$)\footnote{Spearman's $\rho$ is equivalent to Pearson's $r$ on ranked quantities. We note also that $\rho = 0$ does not necessarily indicate a  relationship does not exist; simply that the relationship is not monotonic. A parabolic function for example would result in $\rho = 0$.}. When $\abs{\rho}$ is nearly one, the information from one parameter can be used to determine information about the other. Therefore, this is a valuable tool for probing the relationships that exist in and across evolutionary tracks and determining which model properties can be inferred from which observable quantities. 

In the current analysis, we are strictly interested in the relationships expected from the observational data. We apply cuts to the grid computed by BA1 as it spans a wide parameter range\footnote{When training a RF for the purposes of characterising stellar systems,  sampling the parameter space well beyond the expected ranges of each quantity is prudent. RFs do not extrapolate---doing so would be undesirable anyway---so characterizing a star requires that all of its observations are firmly within the boundaries of the grid used to train the RF. Doing this furthermore avoids pre-conceived biases in the analysis: it allows the observations to dictate the interesting regions of the parameter space rather than limiting the ranges to the values we \emph{expect} the parameters to take.}. The full set of tracks in the BA1 grid includes models with temperatures exceeding the limit in which solar-like oscillations are thought to develop ($ T_{\rm{eff}} \approx 6700$\,K, i.e., the approximate surface temperature beyond which the stellar envelopes are radiative rather than convective). 
Evolutionary tracks in the training grid with more than half of the constituent models having T$_{\rm{eff}} > 6700$\,K are excluded from the rank correlation analysis.
Note that the grid will still contain models with T$_{\rm{eff}} > 6700$\,K if more than half the models in a track display temperatures below this cutoff; there is some chance we may observe such stars. 
Likewise, we omit tracks where high atomic-diffusion rates significantly drain metals from the surface, i.e., tracks where more than half the models display surface-hydrogen mass fractions $> 0.95$. The dearth of stars observed at zero metallicity indicates that there are some physical processes not included in our models (e.g., radiative levitation or turbulent diffusion) which inhibit the unabated flow of metals from the stellar surface.
This is a common result in models of high-mass stars that include gravitational settling and therefore the process is \emph{ordinarily} suppressed once  $M \gtrsim 1.1$\,M$_{\odot}$. Metal depletion may also arise in cases when settling is made to operate extremely efficiently. 
The removal of these sequences reduces the BA1 training set from 5325 to 2010 evolutionary tracks (truncated grid hereinafter) for the current analysis. 
In Figure \ref{fig:HRDRad} we plot the truncated grid in the HR-diagram and color the models according to radius.

Figure \ref{fig:filt_corr} shows the results of the correlation analysis for the truncated grid.
We defer correlation analysis on the full grid of models to Appendix \ref{sec:fullcorr}. 
Care is needed when interpreting Figure \ref{fig:filt_corr}. 
First, it is important to remember that correlation is not transitive\footnote{This is irrespective of whether one is using Pearson's $r$, Spearman's $\rho$ or Kendall's $\tau$.} \citep{lang}, i.e.,
\begin{equation}
\corr(X,Y) \wedge \corr(Y,Z) \not \Rightarrow \corr(X,Z)
\end{equation}
even when the correlations are due to causative relationships \citep{stav}.
In fact one can only draw inference on the direction of $\corr(X,Z)$ in cases when 
\begin{equation}
\rho_{X,Y}^2 + \rho_{Y,Z}^2 > 1
\end{equation}
(transitive criterion hereinafter).

Second, recall that these correlations hold only for the main sequence. During the main sequence there is generally a positive correlation between, say, $L$ and $T_{\rm{eff}}$. 
This relationship will change as the stars evolve further beyond the main-sequence turnoff. 

Third, save for correlations with $\tau_{\rm{MS}}$, the relationships presented here do not describe how parameters correlate internally throughout an evolutionary track. Rather, they describe how they correlate across \emph{all} tracks. For example, as a star ascends the main sequence, luminosity increases and therefore one may expect a strong positive correlation between $\tau$ and $L$. The fact that we report a negative correlation is because higher-mass stars are shorter lived -- thus high $L$ corresponds to a lower $\tau$  when the whole parameter space is considered. This correlation is in fact stronger in the analysis of the complete grid used in BA1 which we report in Appendix \ref{sec:fullcorr}, as our grid truncation preferentially selects against higher-mass stars. 
Furthermore we note that some initial model variables ($M$, Y$_0$, Z$_0$, $\alpha_{\rm{MLT}}$, $\alpha_{\rm{ov}}$ and $D$; all indicated in green) correlate with other parameters.  This would not be the case if we reported correlations within tracks, as these parameters do not change within a given track. 

It should be noted that there is some bias present in the grid as the low-mass stars are not computed to the end of their main-sequence lifetime. The strengths of some correlations would change had we considered evolution beyond the age of the Universe. 
\subsection{Interpreting the Correlations}
Having set the general context in which to interpret Figure \ref{fig:filt_corr}, we highlight some statistical features of the lower main sequence that can be extracted: 

\begin{itemize}
\item Most pairs of parameters with $| \rho | \approx 1$ correspond to well known main-sequence and/or asteroseismic relations. Pairs displaying strong correlations include:

\begin{math}
\langle\Delta\nu_0\rangle  -  \log \rm{g}; \qquad   \langle\Delta\nu_0\rangle - \nu_{\rm{max}}; \qquad     \log \rm{g}  - \nu_{\rm{max}};  \\
\langle\Delta\nu_0\rangle  -  R;   \qquad   \qquad     \log \rm{g} - R;            \qquad              \qquad   M  -  R;  \\
 L - R; \qquad \quad  \qquad\langle\delta\nu_{02}\rangle - X_c.
\end{math}

\item Figure \ref{fig:HRDRad} illustrates why $T_{\rm{eff}}$ and its correlations with $R$ and $L$ are weaker than those listed above. 
Many of the tracks evolve past the main-sequence turn off before exhausting their core-hydrogen abundance. The change in morphology of the HR-diagram and resultant increase in radius impacts on the monotonicity of the respective correlations.

\item The mass of the convective core (M$_{\rm{cc}}$) displays a moderate negative correlation with age whereas it barely registers a relationship with $\tau_{\rm{MS}}$. 
It is the higher-mass and hence shorter-lived stars that preferentially develop convective cores. A negative correlation with age is therefore according to expectations.  
In stars that burn hydrogen radiatively no correlation will develop between M$_{\rm{cc}}$ and $\tau_{\rm{MS}}$. 
In those stars that burn convectively, the size of the convective core will grow but then recede as the CNO-burning region becomes more centrally condensed. 
These two factors lead to an (essentially) null result between M$_{\rm{cc}}$ and $\tau_{\rm{MS}}$.

\item The correlations between $\tau$ and the ratios $\langle r_{02}\rangle$ and $\langle r_{13}\rangle$ are stronger than the correlation between $\tau$ and  X$_c$. 
The grid comprises large ranges in mass and metallicity and hence stars at different ages can possess the same X$_c$, thereby weakening the strength of that correlation.
Conversely, as one might expect,  $\tau_{\rm{MS}}$ exhibits a stronger relationship with X$_c$ than the ratios. 

\item The small frequency separations and the asteroseismic frequency ratios strongly correlate with both $\tau$ and X$_c$. The large frequency separation, however, demonstrates a much stronger correlation with X$_c$ than it does with $\tau$. The rate at which stars burn their central fuel will largely depend on their mass, thus the 
models can attain the same density (which is proportional to the large frequency separation) at a range of ages. Both $\tau_{\rm{MS}}$ and X$_c$ are evolutionary variables and display the expected correlations with $\langle\Delta\nu_0\rangle$. 

\item We lack the necessary information to constrain some of the initial model variables. Indeed $[\rm{Fe/H}]$ provides some constraints on the diffusion efficiency factor $D$, but there is much degeneracy: a model can attain the same surface Y starting with a low $[\rm{Fe/H}]$ and low diffusion rate as a track with a high $[\rm{Fe/H}]$ and high diffusion rate. It is possible that fitting for the base of the convective envelope through seismic analysis of the acoustic glitch signal \citep{2014ApJ...782...18M, 2014ApJ...794..114V} could help further constrain these parameters.

\end{itemize}

Figure \ref{fig:filt_corr} immediately reveals information about the relationships utilised in the machine learning algorithms. 
For those parameter pairs that failed the significance test, neither is likely to feature in the regression model that predicts the other, except in a circumstance where a subset of models exhibit a trend that is absent from the general case of all the models being considered together.
Conversely, where possible, the regressor will attempt to draw on information from pairs that display the strongest correlations.
Quantities such as radius illustrate that there is indeed redundant information in independently measured parameters.
This is useful as the observables measured, and their corresponding accuracy, will vary from survey to survey. 
If a key piece of datum is missing or unreliable, a new regression model can be trained using an appropriate substituted quantity in its place.   
This requires that the redundant information in the observables are treated correctly, if however they are not, then they will lead to biases in model finding procedures. We explore this point further in the next section.

\section{Principle Component Analysis}
\label{sec:PCA}
\begin{figure}
    \centering
    \includegraphics[width=\columnwidth]{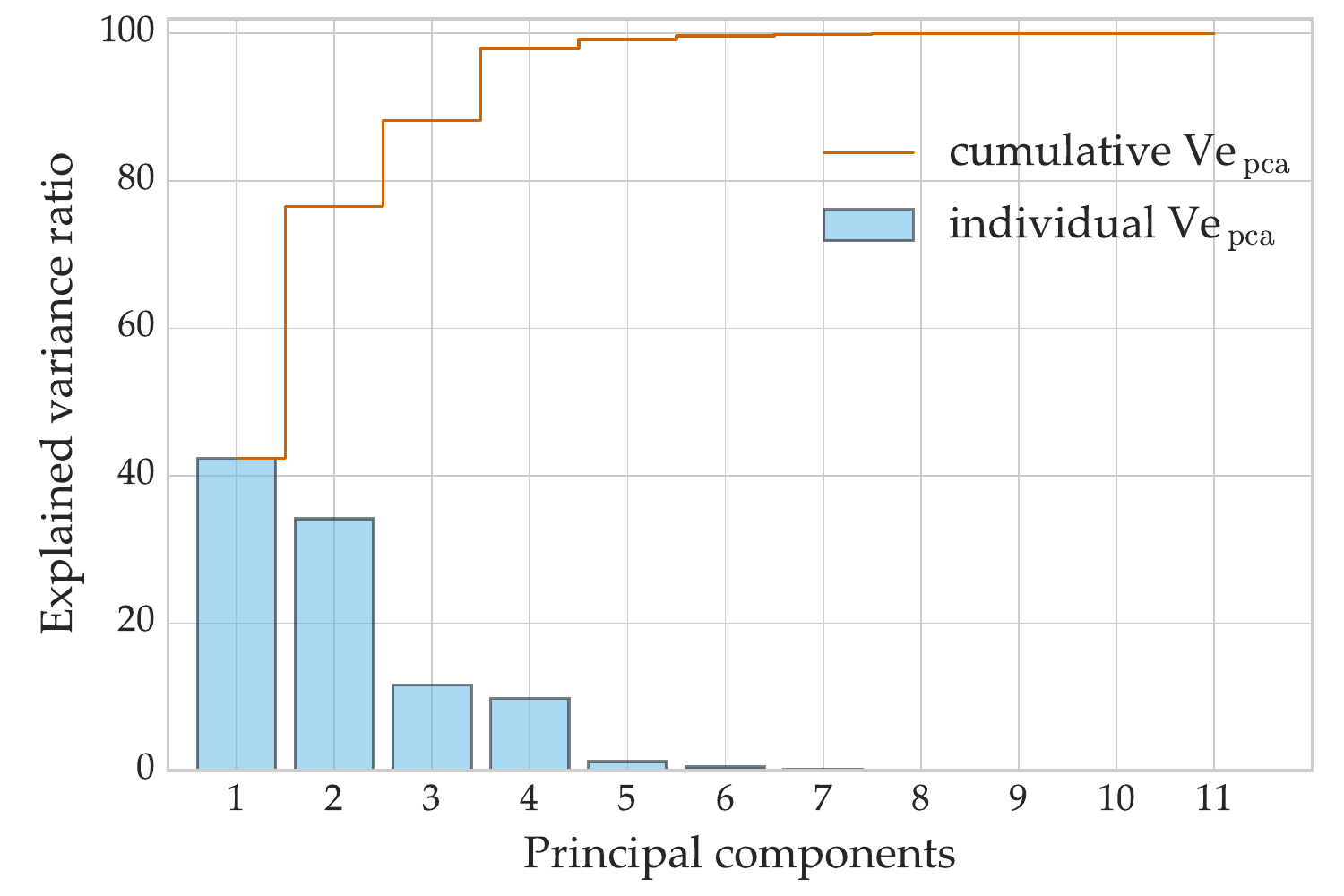}
    \caption{{The explained variance (Ve$_{\,\rm{pca}}$) and cumulative Ve$_{\,\rm{pca}}$ of the principal components comprising the \emph{observable} quantities listed in Table \ref{tab:parmdefs}. The figure demonstrates that 98\% of the variance in the 11 observational parameters can be explained by four independent components and 99.2\% of the variance explained when a fifth component is considered. The Ve$_{\,\rm{pca}}$ of each component is  also presented in the second column of Table \ref{tab:PCAEV}. }}
    \label{fig:GCA-pca}
\end{figure}

\begin{figure*}
\gridline{\fig{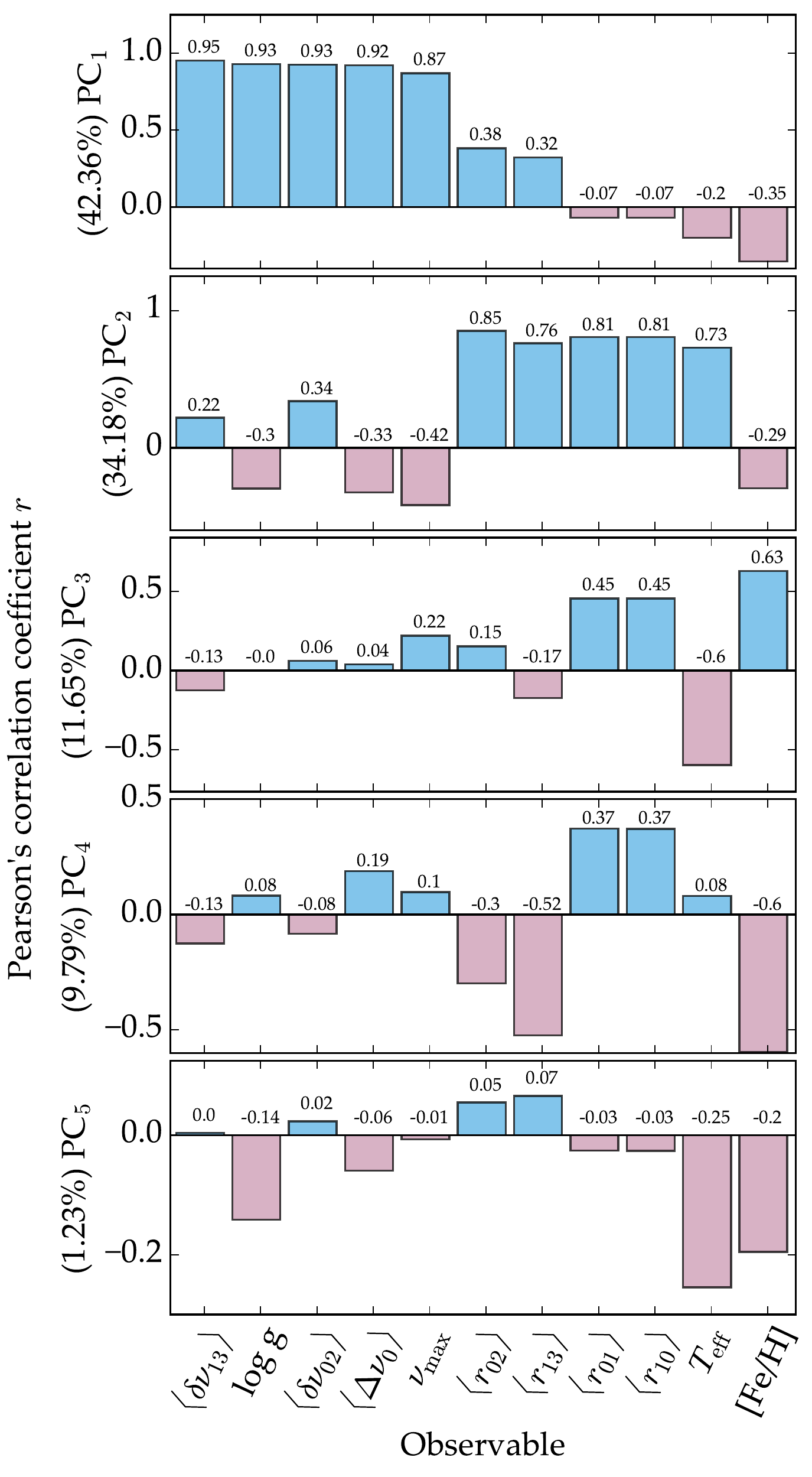}{0.48\textwidth}{\qquad \qquad (a)}
          \fig{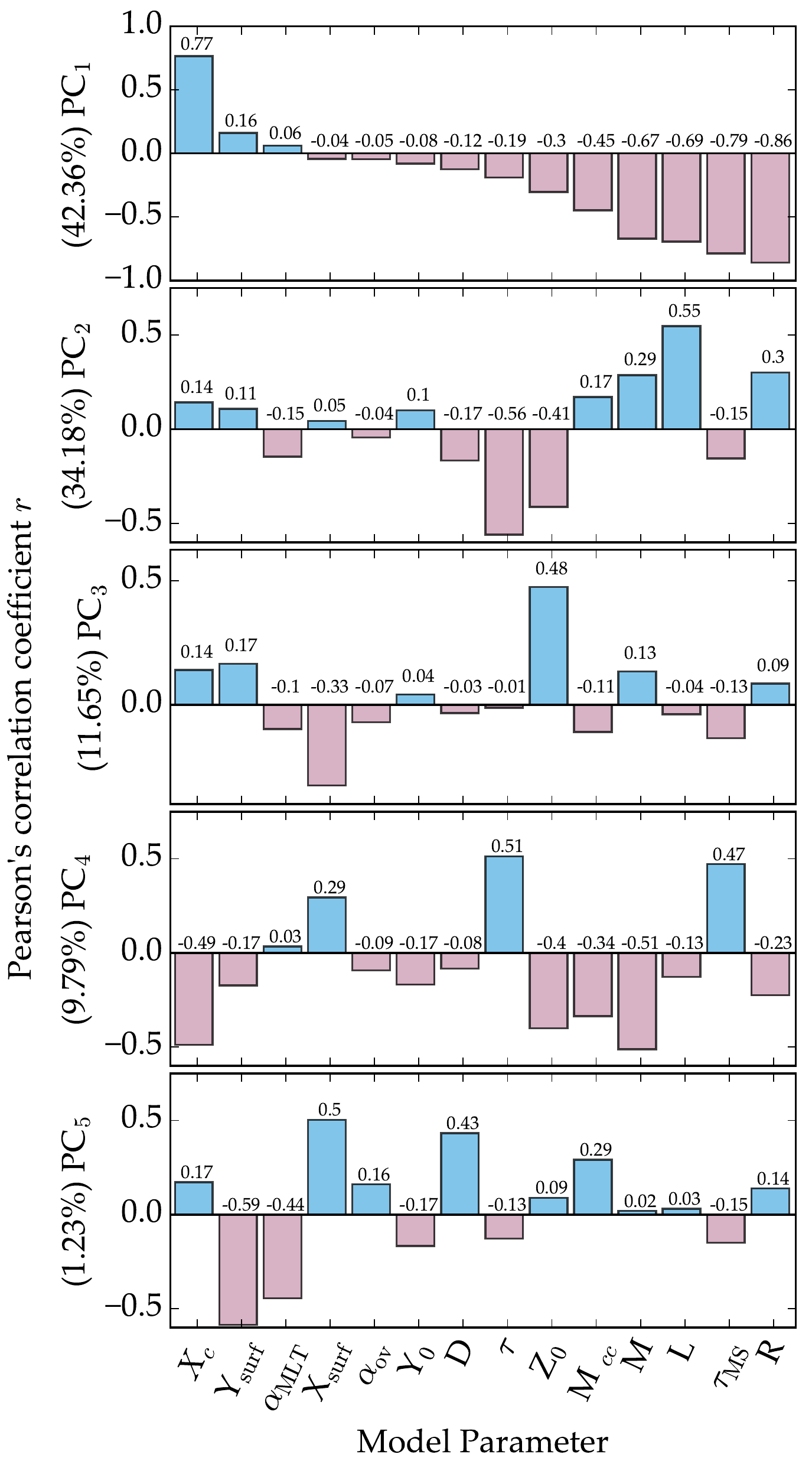}{0.48\textwidth}{\qquad \qquad (b)}
}
\caption{ Pearson correlation strength between the first five principle components and (a) the stellar observables and (b) the model parameters.  Quantities are ordered according to their correlation strength with the first principle component.  Strong correlations indicate that much of the variance of the quantity is captured by the given PC. Note that the ordinate axes in this figure are on different scales. }
\label{fig:GCA-pcabar}
\end{figure*}

Past studies, particularly \citet{Brown1994}, have argued that redundancies and covariances in the stellar observables should be taken into account during any model fitting procedure. 
They demonstrated one particular method (singular value decomposition, SVD hereinafter) of avoiding such biases. In the previous section we identified correlations present in the lower main sequence. Here we demonstrate the degree of redundant information contained in the observables by applying dimensionality reduction. 
We perform principle component analysis (PCA) in order to discover latent structure in observable stellar quantities such that they may be related more directly---and without redundancy---to parameters of stellar modelling. 
Through the principle components (PCs) we quantify the extent to which the observables capture information of the model parameters.

A natural strategy for dealing with high-dimensional data is to reduce the dimensionality in search of \emph{latent variables}; i.e., hidden variables that are more useful than the original quantities under consideration. 
Principle component analysis (PCA) is a technique to transform data into a sequence of orthogonal, and hence independent,  linear combinations of the variables.
Each successive component is constructed to maximize the residual variance from the original data whilst remaining orthogonal to the previous components.
It is a linear transformation in which the change of basis captures the variance contained in original data. 
If parameters in the data are highly correlated, then PCA can potentially produce a lower-dimensional representation without significant loss of the information. 
The method can therefore introduce a new set of variables capable of revealing the underlying structure of an originally high-dimensional space.

PCA  belongs to a family of matrix decomposition techniques that also include methods such as non-negative matrix factorization and independent components analysis as well as variations such as sparse PCA and kernel PCA. 
It has previously been employed in an astrophysical context \citep{2008ApJ...686.1349B, 1987ASSL..131.....M} along with SVD \citep{Brown1994, 2009ApJ...699..373M} to handle correlated errors in observational data. 
The PCs in this work are calculated from the eigensolution of the correlation matrix, the results of which are not scale invariant. 
We note that PCA can be interpreted as the singular value decomposition of a data matrix in cases where the columns have first been centered by their means. 
Thus SVD analysis\footnote{This method is in fact more numerically stable but more computationally expensive for extracting PCs.} is an alternative method for extracting the PCs (see also Appendix \ref{sec:lambdaa}). 
We indeed compare both methods as a check on our methodology and find that the magnitude of PC scores are identical although the direction (sign) of the vector may differ on occasion.

\subsection{Explained Variance of the Principle Components}
\label{sec:ev}
We perform PCA on 11 classical and asteroseismic observables listed in Table \ref{tab:parmdefs}.
The chosen parameters reflect the quantities typically  extracted\footnote{Radius and luminosity are in some cases observable, but not ubiquitously available in the pre-Gaia era. We concede that the inclusion of $\ell=3$ modes is an optimistic assumption.} from stars in the \emph{Kepler} \citep{2004SPIE.5487.1491K,2010Sci...327..977B} field.
Our analysis focuses on the  truncated grid of models\footnote{To extract a robust interpretation of the PCs we consider different subsets of the BA1 grid (see Appendix \ref{sec:fullPCA}).} (see \S \ref{sec:RCT}).
The truncated grid reduces our matrix to size $ 128640 \times 11$ on which we perform the PCA (there are 340800 models in the full BA1 grid). 

The PCs throughout this analysis are calculated from the eigendecomposition of observables in the correlation matrix. Here we wish to explain the variance in the data values rather than their rankings. 
 We employ Pearson's $r$ in the computation of the correlation matrix for the PCA analysis rather than Spearman's $\rho$. This allows us to transform freely back and forth between the original data space and the space of Pearson PCs.  
 
A given data matrix $\mathbf{X}$ (grid) is of size $n \times p$ where $n$ is the number of models  and $p$ is the number of observable parameters. 
Each entry $x_{np}$ in $\mathbf{X}$ is centered and scaled such that 
\begin{equation}
\bar{x}_{np} = (x_{np} -\hat{x_n})/\sigma_{x_n}
\end{equation}
where $\bar{x}_{np}$ is the centered and scaled value,  $x_{np}$ is the original entry, $\hat{x}_n$ is the mean of the particular parameter and $\sigma_{x_n}$ is its standard deviation. 
With all variables having zero mean and unit variance ($\mathbf{\bar{X}}$), our analysis is equivalent to performing eigendecomposition on the covariance matrix\footnote{We are essentially performing the eigendecompostion of the normalized covariance matrix.}. 
We compute $\boldsymbol\Sigma$, the matrix of Pearson's $r$  coefficients, between all entries in $\mathbf{\bar{X}}$; and compute the eigenvalues and eigenvectors of $\boldsymbol\Sigma$ to determine the PCs.
The eigenvalues, $\lambda_i$, of $\boldsymbol\Sigma$ indicate the absolute variance explained by the eigenvectors. We use this to compute the fraction of variance explained by the eigenvector in the dataset, Ve$_{\, \rm{pca}}$, such that:
\begin{equation}
\rm{Ve}_{\, \rm{pca}} \ (\rm{PC}_i)=\frac{\lambda_i}{\sum_{i=1}^p \lambda_i}, 
\label{eqn:pcaev}
\end{equation}
where the number of observables in the data matrix, $p$, is equivalent to the number of principle components we extract.

The Ve$_{\,\rm{pca}}$ and the cumulative explained variance of the PCs are reported in Figure \ref{fig:GCA-pca} (see also the second column in Table \ref{tab:PCAEV}).
Remarkably, we find that 99.2\% of the variance in our 11-dimensional observable space can be explained by a space of five components.
Hence, observable stellar quantities are clearly highly redundant in what they reveal, as only five dimensions contain original information about the star.

Further insight into the PCs can be gained through correlation analysis between the transformed data (i.e., data matrix projected onto the new PC features) and the original data matrix of observables.
Any observable that correlates with a PC contributes to the linear combination of parameters that comprise that PC -- the PC is capturing part of the variance in that observable/dimension. 
Multiple parameters that simultaneously have a large fraction of their variance explained by the same PC, must therefore carry redundant information about the star\footnote{The correlation analysis is in general similar to reporting the PC loadings. 
In PCA loadings are the elements of the eigenvector scaled by the square roots of the respective eigenvalues.
The elements of the eigenvector are coefficients that indicate the weighting of the original data parameters that combine to form that PC. 
As we have centred and scaled the data before performing the PCA the correlation coefficients are equivalent to the loadings.}.
In Figure \ref{fig:GCA-pcabar}a we quantify, through Pearson's $r$ coefficient, the extent to which each observable correlates with the first five PCs.
The parameters in the top panel of Figure \ref{fig:GCA-pcabar}a are ordered by their correlation with the first principle component. 
PC$_1$ accounts for a significant fraction of the variance in the observables (Ve$_{\, \rm{pca}}=42.36\%$). 
The top panel of  Figure  \ref{fig:GCA-pcabar}a reveals that this component correlates very strongly 
($r > 0.85$)  with   
$\nu_{\rm{max}}$, 
$\langle\Delta\nu_0\rangle$,
$\langle\delta\nu_{02}\rangle$, 
$\langle\delta\nu_{13}\rangle$,
and $\log \rm{g}$.  
The strong correlations imply that the basis vector captures most of the variance across the five parameters simultaneously and points to a common latent variable.

\subsection{Interpreting the Principle Components}
\label{sec:intPC}

In Figure  \ref{fig:GCA-pcabar}a and Figure \ref{fig:GCA-pcabar}b we plot the results of correlation analysis between all parameters in the grid and the transformed observables (PCs). 
The figures offer a quantitative overview of the PCs allowing  us to identify what interpretable features the PCs have captured.
We have seen that Figure \ref{fig:GCA-pcabar}a demonstrates the extent to which each observable correlates with the first five PCs, similarly  Figure \ref{fig:GCA-pcabar}b demonstrates how the principle components correlate with the model parameters. The corresponding correlation coefficients between the parameters and \emph{all} PCs are listed in Tables \ref{tab:ocoefs} \& \ref{tab:mcoefs}.

Any interpretation of the PCs based on Figures \ref{fig:GCA-pcabar}a and \ref{fig:GCA-pcabar}b are only valid for the truncated grid of models to which this PCA has been applied. For results on other sub grids we refer the reader to Appendices \ref{sec:fullPCA} and \ref{sec:PCAg}. 
We draw upon the figures for generality in the discussion section (\S \ref{sec:disc}).

Information about \emph{direct} correlations between parameters can be extracted from PCA which further helps with interpreting the underlying features.
Any two parameters that correlate with a given principle component and meet the transitive criterion will be positively correlated if they both have the same sign with respect to the PC, and negatively correlated if their signs differ. 

As is often the case with PCA, the first few principle components can be interpreted as describing the large-scale physical behavior of the system.
We interpret that the underlying feature that PC$_1$ captures is straightforwardly the stellar radius.  
This is the physical property that has the greatest impact on the observables.
From PC$_1$ in Figures \ref{fig:GCA-pcabar}a and \ref{fig:GCA-pcabar}b we can infer (from the transitive criterion) that as a star evolves along the main sequence, i.e., $\tau_{\rm{MS}}$ increases or $X_c$ decreases, radius (and for the most part L) will increase.  
The consequence of increasing radius being $\nu_{\rm{max}}$, 
$\langle\Delta\nu_0\rangle$,
$\langle\delta\nu_{02}\rangle$, 
$\langle\delta\nu_{13}\rangle$,
 $\log \rm{g}$ all decrease and thus their variance is explained by PC$_1$. 
We note that this PC also correlates with $M$ as stars with larger $M$ will have larger radii.

PC$_2$ can be interpreted as a `core-surface' feature.
PC$_2$ correlates strongly with different combinations of seismic ratios and small frequency separations. 
With strong weightings from the core it is no surprise that PC$_2$ features a moderate-to-strong correlation with $\tau$.  
This direction of maximal variance comprises information from all the observables and correlates with (mostly) all the dependent model variables further suggesting some form of time evolution. 
The information from the surface is provided by $T_{\rm{eff}}$.  
There is a degree to which the variance in $T_{\rm{eff}}$ is captured by the time-evolutionary aspect of this component. 
However PC$_2$ also displays a moderate correlation with the time-independent Z$_0$ and thus there is a second aspect to  PC$_2$.
Z$_0$ dictates the temperature at the surface through opacities and nuclear burning in the core.

PC$_3$ appears to have the role of capturing the more-extreme models in the grid. 
In the truncated grid the correlations with $[\rm{Fe/H}]$ and $T_{\rm{eff}}$ suggest that the focus of this PC to account for the variance in the observations imparted by low-metallicity models. 

PC$_4$ appears to be a secondary `core-surface' feature much like PC$_2$.  
It uses surface information, in this case $[\rm{Fe/H}]$, in conjunction with some information from the core in the form of the $\langle r_{02}\rangle$ and $\langle r_{13}\rangle$ ratios.  

PC$_5$ encapsulates the mixing processes that impact upon the surface abundances of the star but it is only required to explain a small fraction of the total variance in the data.

\subsection{Inferring Stellar Parameters} 
\label{sec:ISP}

\begin{table}[]
    \centering
    \begin{tabular}{ccc}
    \hline \hline
   
Parameter &  & $\Lambda_{\rm{param}}$ \\ \hline     
R	&	&	0.97	\\
L	&	&	0.96	\\
$X_c$	&	&	0.94	\\
$\tau_{\rm{MS}}$	&	&	0.93	\\
M	&	&	0.91	\\
$\tau$	&	&	0.79	\\
Z$_0$	&	&	0.73	\\
M$_{cc}$	&	&	0.61	\\
Y$_{\rm{surf}}$	&	&	0.50	\\
X$_{\rm{surf}}$	&	&	0.48	\\
$\alpha_{\rm{MLT}}$	&	&	0.38	\\
Y$_0$	&	&	0.31	\\
D	&	&	0.29	\\
$\alpha_{\rm{ov}}$	&	&	0.08	\\
 \hline
    \end{tabular}
    \caption{The $\Lambda$ score is a sum of the squares of  $r_{\rm{PC, param}}$. Any parameter with high $\Lambda$ is explained well by a linear model and can be confidently inferred. We have insufficient information to constrain those parameters with the lowest $\Lambda$.}
    \label{tab:corrL}
\end{table}

The dimensionality reduction achieved by the PCA quantifies the degree of redundancy in the stellar observables alluded to by Figure \ref{fig:filt_corr}. 
However, we also wish to quantify the extent to which the observed stellar properties constrain the internal structures and chemical mixtures of the star, i.e., the model properties.

In our application of RF regression the machine tries to fit for each model parameter, the success of which we can appraise (see \S \ref{sec:qu}). 
Here we conduct a more fundamental evaluation: how well can we capture the variance in the model parameters simply by explaining the variance in the observed data? 
In other words: having removed the redundancies, to what extent is information of the model parameters encoded in the observables? We hence devise a score, $\Lambda$, such that:
\begin{equation}
   \Lambda(X) = \  \sum^{p}_{i=1}  r(X, PC_i)^2
\end{equation}
where $X$ is the parameter of interest, $p$ is the number of PCs (11 in our case)  and  $r(X, PC_i)$ is the Pearson coefficient between the parameter and the PC.
As we centred and scaled our data before computing the correlation matrix and extracting the PCs, 
 the $\Lambda(X)$ score is equivalent to summating the square of the PC loadings. The square of each loading indicates the variation in an observable that is explained by the component. A useful property of having scaled our data is that $\Lambda(X) =1$  for each of our observables. We demonstrate these properties further in Appendix \ref{sec:lambdaa}.

In Figure \ref{fig:GCA-pcabar}b we projected the parameter space of our model quantities onto the PC space.  Whilst these are not the optimum vectors to explain our model parameters, that is not their purpose -- we instead wish to determine what we can learn about the model quantities by understanding the observables. 
As the square of the correlation coefficients (loadings) will indicate the fraction of explained variance for the parameter by a given PC, determining the  
$\Lambda(X)$ score for the model parameters gives an indication of the extent the model data are retrievable from the observables.  

In Table \ref{tab:corrL} we list the $\Lambda$ score for each of the model parameters in Table \ref{tab:parmdefs}. Parameters with larger $\Lambda$ scores have much of their variance captured by the linear models used to explain the observables.  We expect  to be able to infer parameters such as $R$, $L$ and $\tau_{\rm{MS}}$ with a great deal of confidence through regression. 
Parameters with intermediate values of $\Lambda$ ($\tau$, M$_{cc}$) we can expect to recover with some success by employing more sophisticated modelling, however, it is not clear that there is enough information contained in the observables to always do so. In cases with the lowest values of $\Lambda$, such as the initial model parameters $\alpha_{\rm{MLT}}$, Y$_0$ and $\alpha_{\rm{ov}}$, explaining the variance in the observables does not explain the variance in the model parameters. New observables, that provide independent information about the star, are required to recover these parameters with higher confidence. Fitting the acoustic glitch for example may (eventually) provide constraints on the degree of convective envelope overshoot or atomic diffusion \citep{2017ApJ...837...47V}.

\section{Quantifying the Utility of Stellar Observables} 
\label{sec:qu}

There is certainly value and a degree of intuition in dimensionality reduction.
PCA has demonstrated the significant information redundancy in our data. 
It has also allowed us to identify information from the model parameters manifested in the observables, and indicated to what extent those parameters can be extracted.  
We now turn to another strategy of exploratory data science which is to let machine learning algorithms fit complicated models to the data. As we shift our focus from \emph{what} information is present to \emph{how} it can be exploited, we transition from unsupervised to supervised learning methods.

In the PCA we determined orthogonal vectors that are the best fit to the observables. 
Here we utlize a RF to perform non-parametric, multiple regression in order to create the best functions capable of inferring each stellar parameter. With this particular form of supervised learning 
the relationships between observables and model parameters remain hidden. Some insight into the regression function can be gained through examination of the feature importances but the tree topology makes further interpretation difficult. We thus seek to elucidate the RF's decision making processes by apprasing
 how well different combinations of parameters can predict the quantities in Table \ref{tab:parmdefs}. 
 
 This approach not only illustrates the RF's ability to recover missing observational data, say for a rapid stellar evolution calculation, but also systematically \emph{quantifies} the usefulness of each parameter in predicting all other quantities in the limit of perfect information. It is analogous to a seismic inversion in that it demonstrates the inherent uncertainty with which information can be reconstructed from the available observations. 
Whereas PCA serves to remove the redundant stellar information in the parameters, the analysis here is designed to highlight them. 

Using the full grid of BA1 models, we perform multiple regression on every unique combination of observables in Table \ref{tab:parmdefs}. We omit those combinations that contain the quantity we are training for and include models with $R$ and $L$ as observables, resulting in the calculation of 49\,153 RFs.

We divide the full grid into a testing ($\approx$ 15\,000 models) and training set as per the method ascribed in Appendix D of BA1 so not to over-estimate the performance of the regression. 
We perform two-fold cross-validation on each RF and, as in BA1, measure their success on the test data with an explained variance score, V$_{\text{e}}$:
\begin{equation}
\label{eqn:ve}
  \text{V}_{\text{e}} = 1 - \frac{\text{Var}\{ y - \hat y \}}{\text{Var}\{ y \}}.
\end{equation}
Here $y$ is the \emph{true} value we want to predict, $\hat{y}$ is the predicted value from the random forest, and Var is the variance. This score tells us the extent to which the regressor has reduced the variance in the parameter it is predicting with a score of one implying that the model predicts all values with zero error. This is a different but equivalent definition by which to measure the same quantity in Equation \ref{eqn:pcaev}. We have adopted the same notation as BA1 for evaluating the RF which we use to distinguish from the definition used in the PCA (\S \ref{sec:PCA}). We also provide a measure of the `typical' error in the predictions, $\mu (\epsilon)$,  which is calculated by averaging the absolute difference ($\epsilon$) between the predicted and true values for each parameter. More formally: 
\begin{equation}
\label{eqn:mu}
    \mu (\epsilon) = \frac{1}{n} \sum^n_{i=1} | \hat y_i - y_i |,
\end{equation}
where $n$ is the number of models in the test data. Through $\mu (\epsilon)$  we provide an indicative error associated with the regression model, over the whole parameter space, and  in units of the quantity of interest.  

The best combinations of parameters for inferring each quantity of interest are listed in Table \ref{tab:regparms}. 
We present combinations of up to five parameters after which there is negligible improvement to the predictions. 
We mark with a dash the occasions where the regressor is unable to produce a positive $V_e$ score.  
It is important to remember that while a score of one implies a perfect predictor, any $V_e < 1$ implies there is still \emph{some} error in the model.  We thus opt for truncation rather than rounding when listing the scores. 
Predictions of the seismic quantities are omitted here.  They strongly co-vary and are easily recovered when other seismic parameters are known; they are discussed separately in  \S \ref{sec:seispr}. 
Their strong covariances also mean that many of the ratios and separations used in the regression models are interchangeable (e.g., $\langle r_{02}\rangle$ for 
$\langle r_{13}\rangle$ or $\langle r_{01}\rangle$ for $\langle r_{10}\rangle$) resulting in negligible differences to our two scores. 

Many of the RFs we trained do not provide a satisfactory regression model for the quantity we are training for. Below we provide a deeper analysis for some of the more interesting results, focusing primarily on the predictions of ages and surface abundances. 
\clearpage
\vspace*{5.5cm}
\hspace*{-1.25cm}
\begin{rotatetable}
\begin{deluxetable*}{  l | l  l  l | l  l  l  l | l  l  l  l | l  l  l  l | l  l  l  l   }
\tablewidth{700pt}
\tabletypesize{\scriptsize}
\tablecaption{The best combinations of observables for constraining the non-seismic parameters in Table \ref{tab:parmdefs}. For each combination we provide the $V_e$ score (Equation \ref{eqn:ve}) and $\mu (\epsilon)$ score (Equation \ref{eqn:mu}, given in the units indicated by the predicted quantity column).\label{tab:regparms}}
\tablehead{
           \colhead{Predicted}                                    &
           \multicolumn{3}{c}{One Parameter}               &
           \multicolumn{4}{c}{Two Parameters}               &
           \multicolumn{4}{c}{Three Parameters}               &
           \multicolumn{4}{c}{Four Parameters}               &
           \multicolumn{4}{c}{Five Parameters}        \\        
           \colhead{Quantity}                                          &        
           \colhead{Observable}    &  
           \colhead{$V_e$}    & 
           \colhead{$\mu (\epsilon)$}    & 
           \multicolumn{2}{c}{Observables} &
           \colhead{$V_e$}    & 
           \colhead{$\mu (\epsilon)$}    & 
            \multicolumn{2}{c}{Observables} &
           \colhead{$V_e$}    & 
           \colhead{$\mu (\epsilon)$}    & 
            \multicolumn{2}{c}{Observables} &
           \colhead{$V_e$}    & 
           \colhead{$\mu (\epsilon)$}    & 
            \multicolumn{2}{c}{Observables} &
           \colhead{$V_e$}    & 
           \colhead{$\mu (\epsilon)$}     
           }
\startdata
  R  (R$\sun$)& $\langle\Delta\nu_0\rangle$ & 0.955 & 0.046 & $\langle\Delta\nu_0\rangle$, $\nu_{\rm{max}}$ && 0.985 & 0.027 & $\langle\Delta\nu_0\rangle$, $\nu_{\rm{max}}$, && 0.999 & 0.009 & $\langle\Delta\nu_0\rangle$, $\nu_{\rm{max}}$, && 0.999 & 0.008 & $\langle\Delta\nu_0\rangle$, $\nu_{\rm{max}}$, $T_{\rm{eff}}$, && 0.999 & 0.008\\
  &  &  &  &  &  &  &  &  $T_{\rm{eff}}$ &  &  &  & $T_{\rm{eff}}$, $\log \rm{g}$ &&  & & $\log \rm{g}$, $\langle r_{10}\rangle$ &  &  &\\[3pt]
  $\log \rm{g}$ & $\langle\Delta\nu_0\rangle$ & 0.86 & 0.046 & $T_{\rm{eff}}$, $\nu_{\rm{max}}$ && 0.999 & 0.004 & $T_{\rm{eff}}$, $\nu_{\rm{max}}$, && 0.999 & 0.003 & $T_{\rm{eff}}$, $\nu_{\rm{max}}$, && 0.999 & 0.002 & $T_{\rm{eff}}$, $\nu_{\rm{max}}$, $[\rm{Fe/H}]$, && 0.999 & 0.002\\
  &  &  &  &  &  &  &  &  $[\rm{Fe/H}]$ &  &  &  & $[\rm{Fe/H}]$, $\langle r_{13}\rangle$ & & & & $\langle r_{02}\rangle$, $\langle r_{13}\rangle$ &  &  &\\[3pt]
  L (L$\sun$) & $T_{\rm{eff}}$ & 0.739 & 1.583 & $T_{\rm{eff}}$, $\langle\Delta\nu_0\rangle$ && 0.993 & 0.254 & $T_{\rm{eff}}$, $\langle\Delta\nu_0\rangle$, &&0.999 & 0.13 & $T_{\rm{eff}}$, $\langle\Delta\nu_0\rangle$, && 0.999 & 0.136 & $T_{\rm{eff}}$, $\langle\Delta\nu_0\rangle$, $\nu_{\rm{max}}$, && 0.999 & 0.135\\
  &  &  &  &  &  &  &  &  $\nu_{\rm{max}}$ &  &  &  & $\nu_{\rm{max}}$, $\langle r_{10}\rangle$ &&  & & $\log \rm{g}$, $\langle r_{10}\rangle$ &  &  &\\[3pt]   
 $T_{\rm{eff}}$ (K)& $[\rm{Fe/H}]$ & 0.298 & 1216 & $\log \rm{g}$, $\nu_{\rm{max}}$ && 0.989 & 104 & $\log \rm{g}$, $\nu_{\rm{max}}$, && 0.991 & 95 & $\log \rm{g}$, $\nu_{\rm{max}}$, && 0.992 & 96 & $\log \rm{g}$, $\nu_{\rm{max}}$, $\langle r_{01}\rangle$, && 0.992 & 96\\
  &  &  &  &  &  &  &  &  $\langle r_{01}\rangle$ &  &  &  & $\langle r_{01}\rangle$, $\langle\delta\nu_{13}\rangle$ &&  & & $\langle\Delta\nu_0\rangle$, $\langle\delta\nu_{13}\rangle$ &  &  &\\[3pt]
    Z$_0$ & $[\rm{Fe/H}]$ & 0.927 & 0.003 & $[\rm{Fe/H}]$, $\langle\delta\nu_{02}\rangle$ && 0.96 & 0.002 & $[\rm{Fe/H}]$, $T_{\rm{eff}}$, && 0.982 & 0.001 & $[\rm{Fe/H}]$, $T_{\rm{eff}}$, && 0.986 & 0.001 & $[\rm{Fe/H}]$, $T_{\rm{eff}}$, $\langle\Delta\nu_0\rangle$, && 0.987 & 0.001\\
  &  &  &  &  &  &  &  &  $\langle\Delta\nu_0\rangle$ &  &  &  & $\langle\Delta\nu_0\rangle$, $\langle r_{13}\rangle$ &&  & & $\langle r_{01}\rangle$, $\langle r_{13}\rangle$ &  &  &\\[3pt]
  M (M$\sun$)& $\langle\Delta\nu_0\rangle$ & 0.348 & 0.157 & $\langle\Delta\nu_0\rangle$, $\log \rm{g}$ && 0.857 & 0.072 & $\langle\Delta\nu_0\rangle$, $T_{\rm{eff}}$, && 0.982 & 0.022 & $\langle\Delta\nu_0\rangle$, $\log \rm{g}$, && 0.986 & 0.02 & $\langle\Delta\nu_0\rangle$, $\log \rm{g}$, $\nu_{\rm{max}}$, && 0.982 & 0.024\\
  &  &  &  &  &  &  &  &  $\nu_{\rm{max}}$ &  &  &  & $\nu_{\rm{max}}$, $T_{\rm{eff}}$ &&  & & $T_{\rm{eff}}$, $\langle r_{10}\rangle$ &  &  &\\[3pt] 
    $\tau_{\rm{MS}}$ & $\langle\delta\nu_{02}\rangle$ & 0.543 & 0.147 & $\langle r_{02}\rangle$, $\langle r_{01}\rangle$ && 0.846 & 0.077 & $\langle\Delta\nu_0\rangle$, $\nu_{\rm{max}}$, && 0.957 & 0.038 & $\langle r_{02}\rangle$, $\nu_{\rm{max}}$, && 0.977 & 0.025 & $\langle r_{02}\rangle$, $\nu_{\rm{max}}$, $\langle r_{10}\rangle$, && 0.981 & 0.021\\
  &  &  &  &  &  &  &  &  $\langle r_{13}\rangle$ &  &  &  & $\langle r_{10}\rangle$, $T_{\rm{eff}}$ &&  & & $T_{\rm{eff}}$, $[\rm{Fe/H}]$ &  &  &\\[3pt] 
   X$_c$ & $\langle\delta\nu_{02}\rangle$ & 0.508 & 0.113 & $\nu_{\rm{max}}$, $\langle r_{13}\rangle$ && 0.842 & 0.062 & $\nu_{\rm{max}}$, $\langle r_{13}\rangle$, && 0.958 & 0.031 & $\nu_{\rm{max}}$,  $\langle r_{13}\rangle$,& & 0.978 & 0.023 & $\nu_{\rm{max}}$, $\langle r_{13}\rangle$, $\langle\Delta\nu_0\rangle$, && 0.979 & 0.022\\
  &  &  &  &  &  &  &  &  $\langle\Delta\nu_0\rangle$ &  &  &  & $\langle\Delta\nu_0\rangle$, $\langle r_{10}\rangle$ &&  & & $\log \rm{g}$, $\langle r_{10}\rangle$ &  &  &\\[3pt]
  $\tau$ (Gyr)& $\langle r_{02}\rangle$ & 0.645 & 0.995 & $\langle r_{02}\rangle$, $\nu_{\rm{max}}$ && 0.844 & 0.642 & $\langle r_{13}\rangle$, $\nu_{\rm{max}}$, && 0.907 & 0.468 & $\langle r_{02}\rangle$,  $T_{\rm{eff}}$, && 0.931 & 0.332 & $\langle r_{02}\rangle$, $\nu_{\rm{max}}$, $\langle r_{01}\rangle$, && 0.943 & 0.282\\
  &  &  &  &  &  &  &  &  $\langle r_{10}\rangle$ &  &  &  & $\langle r_{01}\rangle$, $\langle\Delta\nu_0\rangle$ &&  & & $T_{\rm{eff}}$, $[\rm{Fe/H}]$ &  &  &\\[3pt] 
   X$_{\rm{surf}}$ & $[\rm{Fe/H}]$ & 0.655 & 0.051 & $[\rm{Fe/H}]$, $\log \rm{g}$ && 0.772 & 0.041 & $[\rm{Fe/H}]$, $\langle\Delta\nu_0\rangle$, && 0.895 & 0.027 & $[\rm{Fe/H}]$, $\langle\Delta\nu_0\rangle$, && 0.928 & 0.024 & $[\rm{Fe/H}]$, $\langle\Delta\nu_0\rangle$, $T_{\rm{eff}}$, && 0.936 & 0.022\\
  &  &  &  &  &  &  &  &  $\nu_{\rm{max}}$ &  &  &  & $T_{\rm{eff}}$, $\langle r_{02}\rangle$ &&  & & $\langle r_{02}\rangle$, $\nu_{\rm{max}}$ &  &  &\\[3pt] 
  M$_{\rm{cc}}$ (M$\sun$)& --- & --- & --- & $\langle r_{13}\rangle$, $\langle\delta\nu_{02}\rangle$ && 0.679 & 0.015 & $\langle r_{13}\rangle$, $\nu_{\rm{max}}$, & &0.862 & 0.009 & $\langle r_{13}\rangle$, $\nu_{\rm{max}}$, && 0.908 & 0.007 & $\langle r_{13}\rangle$, $\nu_{\rm{max}}$, $\langle r_{10}\rangle$, && 0.928 & 0.006\\
  &  &  &  &  &  &  &  &  $\langle r_{10}\rangle$ &  &  &  & $\langle r_{10}\rangle$, $T_{\rm{eff}}$ &&  & & $T_{\rm{eff}}$, $[\rm{Fe/H}]$ &  &  &\\[3pt]
   Y$_{\rm{surf}}$ & $[\rm{Fe/H}]$ & 0.597 & 0.052 & $[\rm{Fe/H}]$, $\log \rm{g}$ && 0.736 & 0.041 & $[\rm{Fe/H}]$, $\langle\Delta\nu_0\rangle$, && 0.887 & 0.025 & $[\rm{Fe/H}]$, $\langle\Delta\nu_0\rangle$, && 0.916 & 0.024 & $[\rm{Fe/H}]$, $\langle\Delta\nu_0\rangle$, $\langle r_{02}\rangle$, && 0.927 & 0.022\\
  &  &  &  &  &  &  &  &  $\nu_{\rm{max}}$ &  &  &  & $\langle r_{02}\rangle$, $T_{\rm{eff}}$ &&  & & $T_{\rm{eff}}$, $\nu_{\rm{max}}$ &  &  &\\[3pt] 
   Y$_0$ & --- & --- & --- & $\langle\Delta\nu_0\rangle$, $\nu_{\rm{max}}$ && 0.077 & 0.027 & $\langle\Delta\nu_0\rangle$, $\nu_{\rm{max}}$, && 0.471 & 0.02 & $\langle\Delta\nu_0\rangle$, $\nu_{\rm{max}}$, && 0.536 & 0.019 & $\langle\Delta\nu_0\rangle$, $\nu_{\rm{max}}$, $[\rm{Fe/H}]$, && 0.625 & 0.017\\
  &  &  &  &  &  &  &  &  $[\rm{Fe/H}]$ &  &  &  & $[\rm{Fe/H}]$, $\log \rm{g}$ &&  & & $\log \rm{g}$, $\langle\delta\nu_{13}\rangle$ &  &  &\\[3pt] 
    $\alpha_{\rm{ov}}$ & --- & --- & --- & $\langle r_{13}\rangle$, $\langle r_{02}\rangle$ && 0.231 & 0.089 & $\langle r_{13}\rangle$, $\langle r_{10}\rangle$, && 0.44 & 0.075 & $\langle r_{13}\rangle$,  $\langle r_{10}\rangle$, && 0.524 & 0.068 & $\langle r_{13}\rangle$, $\langle r_{10}\rangle$, $\nu_{\rm{max}}$, && 0.55 & 0.067\\
  &  &  &  &  &  &  &  &  $\nu_{\rm{max}}$ &  &  &  & $\nu_{\rm{max}}$, $T_{\rm{eff}}$ &&  & & $T_{\rm{eff}}$, $[\rm{Fe/H}]$ &  &  &\\[3pt] 
   D & --- & --- & --- & $[\rm{Fe/H}]$, $\langle\delta\nu_{02}\rangle$ && 0.022 & 5.393 & $[\rm{Fe/H}]$, $T_{\rm{eff}}$, && 0.295 & 4.483 & $[\rm{Fe/H}]$, $T_{\rm{eff}}$, && 0.446 & 3.706 & $[\rm{Fe/H}]$,  $T_{\rm{eff}}$, $\langle r_{02}\rangle$, && 0.519 & 3.333\\
  &  &  &  &  &  &  &  &  $\langle\Delta\nu_0\rangle$ &  &  &  & $\langle r_{02}\rangle$, $\langle\Delta\nu_0\rangle$ &&  & & $\log \rm{g}$, $\langle r_{10}\rangle$ &  &  &\\[3pt] 
  $[\rm{Fe/H}]$ &--- & --- & --- & $\nu_{\rm{max}}$, $\log \rm{g}$ && 0.179 & 2.777 & $\nu_{\rm{max}}$, $\log \rm{g}$, && 0.273 & 2.439 & $\nu_{\rm{max}}$, $\log \rm{g}$, && 0.309 & 2.312 & $\nu_{\rm{max}}$, $\log \rm{g}$, $\langle r_{02}\rangle$, && 0.312 & 2.277\\
  &  &  &  &  &  &  &  &  $\langle r_{02}\rangle$ &  &  &  & $\langle r_{02}\rangle$, $\langle r_{10}\rangle$ &&  & & $\langle r_{01}\rangle$, $\langle r_{13}\rangle$ &  &  &\\[3pt] 
  $\alpha_{\rm{MLT}}$ & --- & --- & --- & --- --- && --- & --- & $T_{\rm{eff}}$, $\nu_{\rm{max}}$, && 0.069 & 0.234 & $T_{\rm{eff}}$, $\nu_{\rm{max}}$, && 0.201 & 0.211 & $T_{\rm{eff}}$, $\nu_{\rm{max}}$, $\langle r_{01}\rangle$, && 0.229 & 0.207\\
  &  &  &  &  &  &  &  &  $\langle\delta\nu_{13}\rangle$ &  &  &  & $[\rm{Fe/H}]$, $\langle r_{02}\rangle$ &&  & & $[\rm{Fe/H}]$, $\langle r_{13}\rangle$ &  &  &\\[3pt] 
\enddata
 \end{deluxetable*} 
\end{rotatetable}
\clearpage

\subsection{Ages}
\label{sec:sages}
\begin{table}
\centering
\caption{The best two-parameter combinations of observables for constraining stellar age. Below the dividing horizontal line we include the best spectroscopic pair for comparison as well as $\log \rm{g}$ -- $\langle\Delta\nu_0\rangle$ to highlight the necessity of the small frequency separation in determining stellar ages. The BA1 grid is varied in six dimensions and with such a high-dimensional parameter space the quantities in the C-D diagram (fifth row) constrain age with `typical' uncertainty of 701 Myr.}
    \begin{tabular}{cccc}
    \hline \hline
\multicolumn{2}{c}{Parameters} & $V_e$ & $\mu (\epsilon)$ [Gyr] \\ \hline 
$\langle r_{02}\rangle$     & $\nu_{\rm{max}}$               &0.844  & 0.642\\
$\langle r_{02}\rangle$     & $\log \rm{g}$                  &0.833  & 0.683\\
$\langle r_{13}\rangle$     &$\nu_{\rm{max}}$                &0.827  & 0.711\\
$\langle r_{02}\rangle$     & $\langle\Delta\nu_0\rangle$    &0.825  & 0.694\\
$\langle\Delta\nu_0\rangle$ &$\langle\delta\nu_{02}\rangle$  &0.824  & 0.701\\
$\langle r_{02}\rangle$     & $\langle\delta\nu_{02}\rangle$ &0.821  & 0.701\\
PC$_2$ & PC$_8$ & 0.788 & 0.767 \\
PC$_2$ & PC$_4$ & 0.776 & 0.762 \\
\hline
 $\log \rm{g}$ & $\langle\Delta\nu_0\rangle$ & 0.481 & 1.29 \\
 $\log \rm{g}$ & $T_{\rm{eff}}$ & 0.321 & 1.53 \\
 \hline
    \end{tabular}
    \label{tab:cd}
\end{table}
The current exercise allows us to evaluate the theoretical limit in which parameter pairs, such as those used in the C-D diagram, can constrain stellar ages. 
Recall that there are six initial model parameters varied simultaneously in the BA1 grid. 
Describing a six dimensional parameter space with two quantities invariably leads to degenerate solutions for age and necessarily high uncertainties.
The parameter pairs that offer similarly the best constraints on $\tau$ are listed in Table \ref{tab:cd}. The combination of $\langle r_{02}\rangle$ and $\nu_{\rm{max}}$  marginally provide the best probe, explaining the largest fraction of the variance and inferring ages with uncertainty $\mu (\epsilon) = \pm 642$ Myr.  
This is in comparison to $\mu (\epsilon) = \pm 701$ Myr for $\langle\Delta\nu_0\rangle$ and $\langle\delta\nu_{02}\rangle$  as per the C-D diagram. 
In Table \ref{tab:cd} we also include results from regression calculated with the PCs and find they perform comparably well. The results here omit any uncertainty stemming from the surface effect suggesting that the $\langle r_{02}\rangle$ and $\nu_{\rm{max}}$ pair are indeed the preferable choice.

It is clear from Tables \ref{tab:regparms} and \ref{tab:cd}  how important the small frequency separation and frequency ratios are for the determination of stellar ages on the MS. 
If we limit the combinations to the classical observables, we find that  $\log \rm{g}$ and $T_{\rm{eff}}$ can explain just 32.1\% of the variance in $\tau$ with uncertainty $\mu (\epsilon) = \pm 1.5$ Gyr across the whole grid. The introduction of the large separation offers little improvement. The parameter pair $\log \rm{g}$ and $\langle\Delta\nu_0\rangle$  explain 48.1\% of the variance with  $\mu (\epsilon) = \pm 1.29$ Gyr.
If we permit the RF to draw upon five observables for its regression model, some of the degeneracy in $\tau$ is lifted.  The last column in Table \ref{tab:regparms}
indicates that the RF can reduce the average uncertainty in predicting  $\tau$ such that  $\mu (\epsilon) =\pm 282$ Myr.

\subsection{Abundances}
\begin{figure}[ht]
    \centering
    \includegraphics[width=\columnwidth]{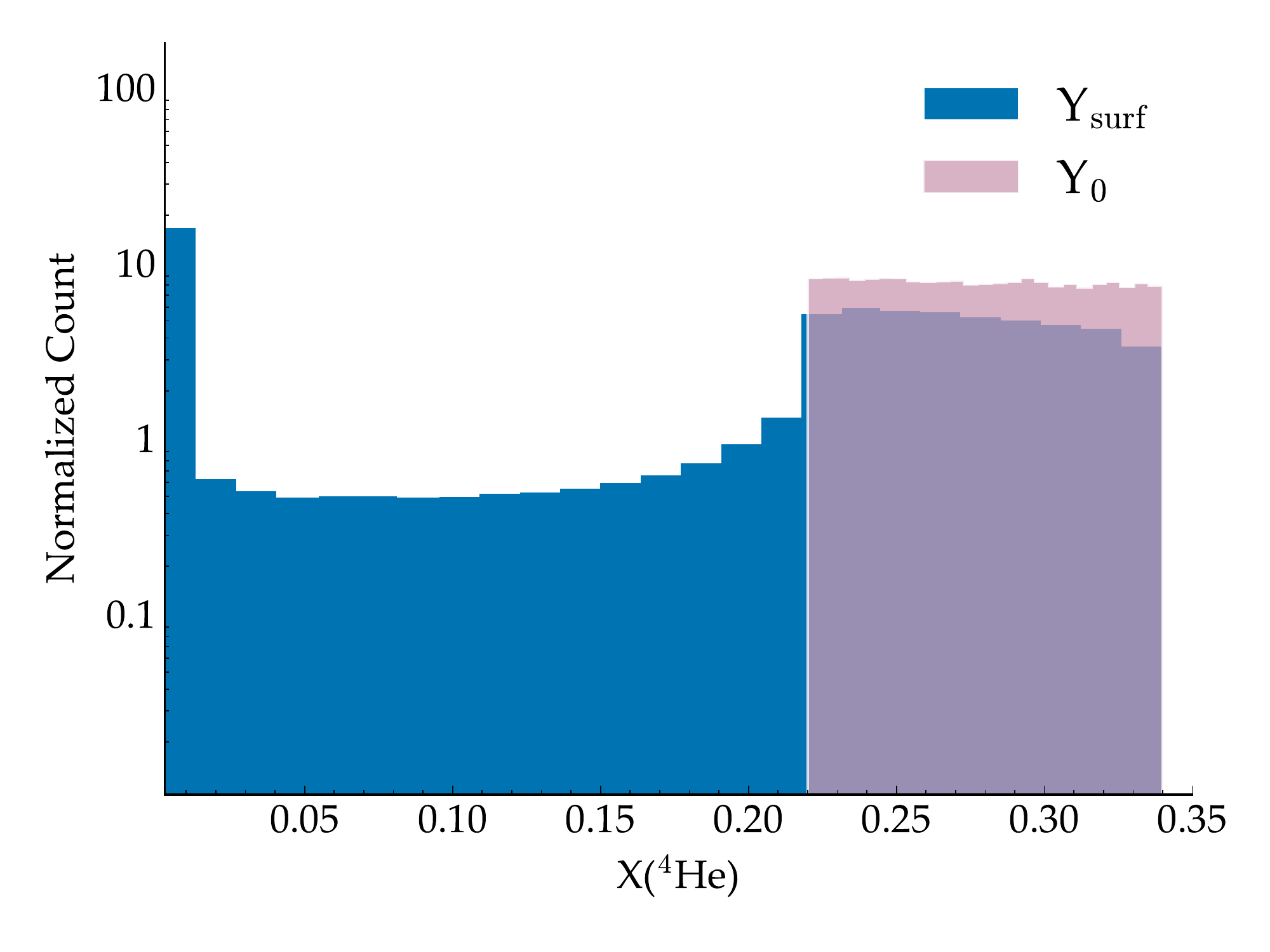}
    \caption{Distributions of $Y_{\rm{surf}}$ and Y$_0$ in the BA1 grid.}
    \label{fig:Hehist}
\end{figure}
The small frequency separations and separation ratios are integral for the determination of ages. 
However, the feature importances in BA1 (their Figure 5) indicate that the RF relies predominately on  $T_{\rm{eff}}$ and $[\rm{Fe/H}]$ to infer other model parameters. 
Table \ref{tab:regparms} confirms how important measuring $[\rm{Fe/H}]$ is for characterizing stars. This quantity is preferentially selected in the many RFs and their regression models, whilst $[\rm{Fe/H}]$ itself cannot be determined from the other observables with any degree of confidence.   
$[\rm{Fe/H}]$ is an indispensable piece of independent information.

Accurate determination of $[\rm{Fe/H}]$ is paramount for inferring many of the current-age stellar attributes. $[\rm{Fe/H}]$ also features prominently in the retrodiction of the initial model parameters but these quantities are characterized by large uncertainties. Foremost, we have no observable that satisfactorily constrains diffusion -- $D$ demonstrates an average uncertainty spanning three orders of magnitude. This in turn introduces uncertainty in retrodicting the initial metal content. 

Predictions for Z$_0$ at first glace appear to be robust; we report $V_e$ and  $\mu (\epsilon) = \pm 0.001$. However we contend that a reported error of  $\mu (\epsilon) = \pm 0.001$ is not all that insightful given that the grid is sampled down to $Z_0 = 10^{-5}$. Z$_0$ is sampled logarithmically and takes a small (linear) range in values. In such cases   a relative error is a more useful measure of performance than an absolute difference.

In Table \ref{tab:relabunds} we devise a series of measures that better appraise the performance of the RF in predicting abundances. We report the average absolute difference as per Table \ref{tab:regparms} [$\mu (\epsilon)$], the maximum absolute difference [Max($\epsilon$)] and the median absolute difference [$\tilde{\epsilon}$]. We also consider the average relative error [$\mu (\eta)$], the maximum relative error [Max($\eta$)] and median relative error [$\tilde{\eta}$], where the relative error is a percentage defined as
\begin{equation}
  \eta= \frac{| \hat y_i - y_i |}{|y_i|} \cdot 100.
\end{equation}

We find  $\mu (\eta) =125\,\%$ in the retrodiction of metallicity. We attribute the seemingly large uncertainty to the bias imparted by extreme models that have undergone significant diffusion -- we report a maximum relative error of 9000\,\%. With less sensitivity to the outlying metal-depleted models, the median relative uncertainty, $\tilde{\eta} = 13.5\,\% $, offers the most appropriate measure of error in the regression. 
Likewise, the extreme $\mu (\eta)$ and Max($\eta$) scores for $Y_{\rm{surf}}$ also stem from models with high diffusion leading to very small non-zero abundances by which we normalise. 

It is interesting to compare the regressor's ability to infer $Y_{\rm{surf}}$ and Y$_0$ abundances. We find that  $Y_{\rm{surf}}$ can be well fit ($V_e = 0.927$) with $\mu (\epsilon) = \pm 0.022$. In contrast, the initial abundance, Y$_0$, cannot be confidently retrodicted  ($V_e$=0.625) yet results in a smaller average error [$\mu (\epsilon) = \pm 0.017$]. 
This initially surprising result can be understood through examination of the respective parameter distributions in the BA1 grid (Figure \ref{fig:Hehist}).
The grid is uniformly sampled in initial helium with $Y_0 \in [0.22, 0.34]$. 
Atomic diffusion acts to drain helium from the surface layers and in fact, in some models, completely depletes this species from the envelope. 
The surface helium abundance of a stellar model can thus attain values in the larger range
$Y_{\rm{surf}} \in [0.0, 0.34]$.  
In a  uniform distribution, such as we have for Y$_0$, the largest theoretical uncertainty  is 
\begin{equation}
    \max \left( \frac{\sigma^2(Y_0)}{Y_0} \right) = \frac{|b-a|}{|a|} \cdot 100 = 54.51\,\%,
\end{equation}
where $a$ and $b$ are the respective minimum and maximum values in our parameter range.  
This means that if the regressor was unable to explain \emph{any} of the variance in this quantity and was randomly choosing Y$_0$ values from the initial distribution, the worst relative uncertainty we would expect is 54.51\,\%. The fact that we do go someway to predicting this quantity results in $\mu (\eta) \approx 8\,\% $ and more accurate inferences than for $Y_{\rm{surf}}$.

\begin{table}[h]
\centering
\caption{Different measures of uncertainty in predicting stellar abundances with the RF. See text for definitions and motivations.}
    \begin{tabular}{lccc}
 \hline \hline
 Error Measure& $Y_{\rm{surf}}$ & Y$_0$ & Z$_0$ \\ \hline
 $\mu (\epsilon)$                 & 0.02 &0.017 & 0.001 \\
 Max($\epsilon$)             & 0.25 &0.09   & 0.037\\    
$\tilde {\epsilon}$          & 0.016  &0.02     &0.00019\\
$\mu (\eta)$ [\%]       & $10^{13}$ &8.92& 124.5 \\ 
  Max($\eta$) [\%]   & $10^{14}$ &40.34  & 9052 \\
$\tilde {\eta}$  [\%] & 10.88 &7.68  & 13.5 \\
\hline
\end{tabular}
\label{tab:relabunds}
\end{table}

\subsection{Other Results}
We mention briefly other interesting results from the approximately 50\,000 RFs not necessarily reported in Table \ref{tab:regparms}.
Stellar masses can be accurately inferred from spectroscopic measurements. The combination of  $\log \rm{g}$, $T_{\rm{eff}}$ and $[\rm{Fe/H}]$ constrains mass equally well as the pair $\langle\Delta\nu_0\rangle$ -- $\log \rm{g}$. Both combinations explain 
 86\,\% of the variance in mass with $\mu (\epsilon) = \pm 0.07$\,M$_{\sun}$. With six degrees of freedom in the BA1 grid, we cannot determine mass to an accuracy better than  $\mu (\epsilon) = \pm 0.02\,\rm{M}_{\sun}$.  
 Whilst all observables correlate with $M$, they do not contain sufficient information to separate out the redundant structures that are possible by tweaking the other initial model parameters. We in fact find no improvement in our regression for $M$ beyond three parameters\footnote{Numerics accounts for the differences in the third decimal place for scores in Table \ref{tab:regparms}.}. 

If required, the RF can determine $T_{\rm{eff}}$ with high accuracy. 
Although this is almost certainly always an input for the RF, with two or more observables $T_{\rm{eff}}$ can be determined with  $\mu (\epsilon) \approx 100$\,K -- an uncertainty comparable to typical spectroscopic errors. If one of $L$ or $R$ are provided as an input to the RF, a factor of two reduction in the uncertainty is achieved with $\mu (\epsilon) \lesssim 50$\,K. Furthermore, our testing of the RF (not included here) indicates that if both $L$ and $R$ are provided as observables the Stefan-Boltzmann law is recovered with $\mu (\epsilon) = 4$\,K.

\subsection{Seismic Quantities} \label{sec:seispr}
We did not include the predictions for the seismic parameters in Table \ref{tab:regparms} as they often carry redundant information. Indeed we accomplish little 
by reporting how the different combinations of ratios and separations can be used to recover each other. 
We thus opt to analyse the seismic parameters separately, where we can employ discretion to present useful comparisons and highlight noteworthy results. 

\subsubsection{The large frequency separation -- $\langle\Delta\nu_0\rangle$}
\begin{table}[h]
\centering
\caption{Combinations of observables that best constrain $\langle\Delta\nu_0\rangle$.}
    \label{tab:dnu}
    \begin{tabular}{ccccccc}
    \hline \hline
\multicolumn{3}{c}{Parameters} && $V_e$ & $\mu (\epsilon)$ & $\mu (\eta)$  \\  
&&&&&[$\mu$Hz] & [\%] \\ \hline
$\nu_{\rm{max}}$ &  &   &&0.930 & 7.815 & 6.11 \\ 
$T_{\rm{eff}}$     & $\nu_{\rm{max}}$&     &&0.990  & 3.09 & 2.46 \\
$\log \rm{g}$     &$\nu_{\rm{max}}$ &               &&0.990  & 2.95 & 2.34 \\
$\log \rm{g}$     &  $T_{\rm{eff}}$&                 & &0.990  & 2.92 & 2.31\\
$[\rm{Fe/H}]$  & $\nu_{\rm{max}}$ &                 &&0.991  & 2.81 & 2.24\\
$T_{\rm{eff}}$ & $[\rm{Fe/H}]$ & $\nu_{\rm{max}}$ && 0.995 & 1.67 & 2.13 \\
$\log \rm{g}$ & $[\rm{Fe/H}]$ & $\nu_{\rm{max}}$ && 0.995 & 1.65 & 2.11\\
\hline
    \end{tabular}
\end{table}

\begin{table}[]
\centering
\caption{Predictions of $\langle\Delta\nu_0\rangle$ for stars listed in \citet{2009MNRAS.400L..80S}. Results pertain to a random forest trained with $\nu_{\rm{max}}$ as the only input. Predictions are compared to literature values from the sources listed in Table 1 of  \citet{2009MNRAS.400L..80S}. The RF performs as well as the power-law relation (10-15\%) even on data measured with less precision than stars observed by \emph{Kepler}.}
\label{tab:stello}
\begin{tabular}{llllll}
\hline  \hline
Star & $\nu_{\rm{max}}$ & $\langle\Delta\nu_0\rangle_{\rm{lit}}$ & $\langle\Delta\nu_0\rangle_{\rm{pred}}$ & $\epsilon$ & $\eta$  \\ 
& ($\mu$Hz) & ($\mu$Hz)  &($\mu$Hz) &($\mu$Hz) & (\%) \\ \hline
$\tau\,$Cet &4500 & 170 & 171 & 1  & 1  \\
$\alpha\,$Cen B &4100 & 161 & 184 & 22& 14  \\
Sun &3100 & 135 & 138 & 3 & 2    \\
$\iota\,$Hor &2700 & 120   & 136 &  16&14  \\
$\gamma\,$Pav  &2600 & 120 & 122 & 1  & 1  \\
$\alpha\,$Cen A &2400 & 106 & 124 & 18 & 17  \\
HD175726 &2000 & 97    & 100 & 3  & 3  \\
$\mu\,$Ara &2000 & 90    & 100 & 10 & 11  \\
HD181906 &1900 & 88  & 97  & 10 & 11 \\
HD49933 &1760 & 86  & 101 & 15 & 18  \\
HD181420 &1500 & 75    & 76  & 1  & 1  \\
$\beta\,$Vir &1400 & 72    & 77   & 5 & 8  \\
$\mu\,$Her  &1200 & 57  & 63   & 7 & 12  \\
$\beta\,$Hyi &1000 & 57  & 57  & 0  & 0  \\
Procyon &1000 & 55    & 57 & 2 & 4    \\
$\eta\,$Boo &750  & 40  & 45   & 5 & 13  \\
$\nu\,$Ind &320  & 25  & 23  & 3 & 10  \\ \hline
\end{tabular}
\end{table}

\begin{figure}
\centering
\includegraphics[width=\columnwidth]{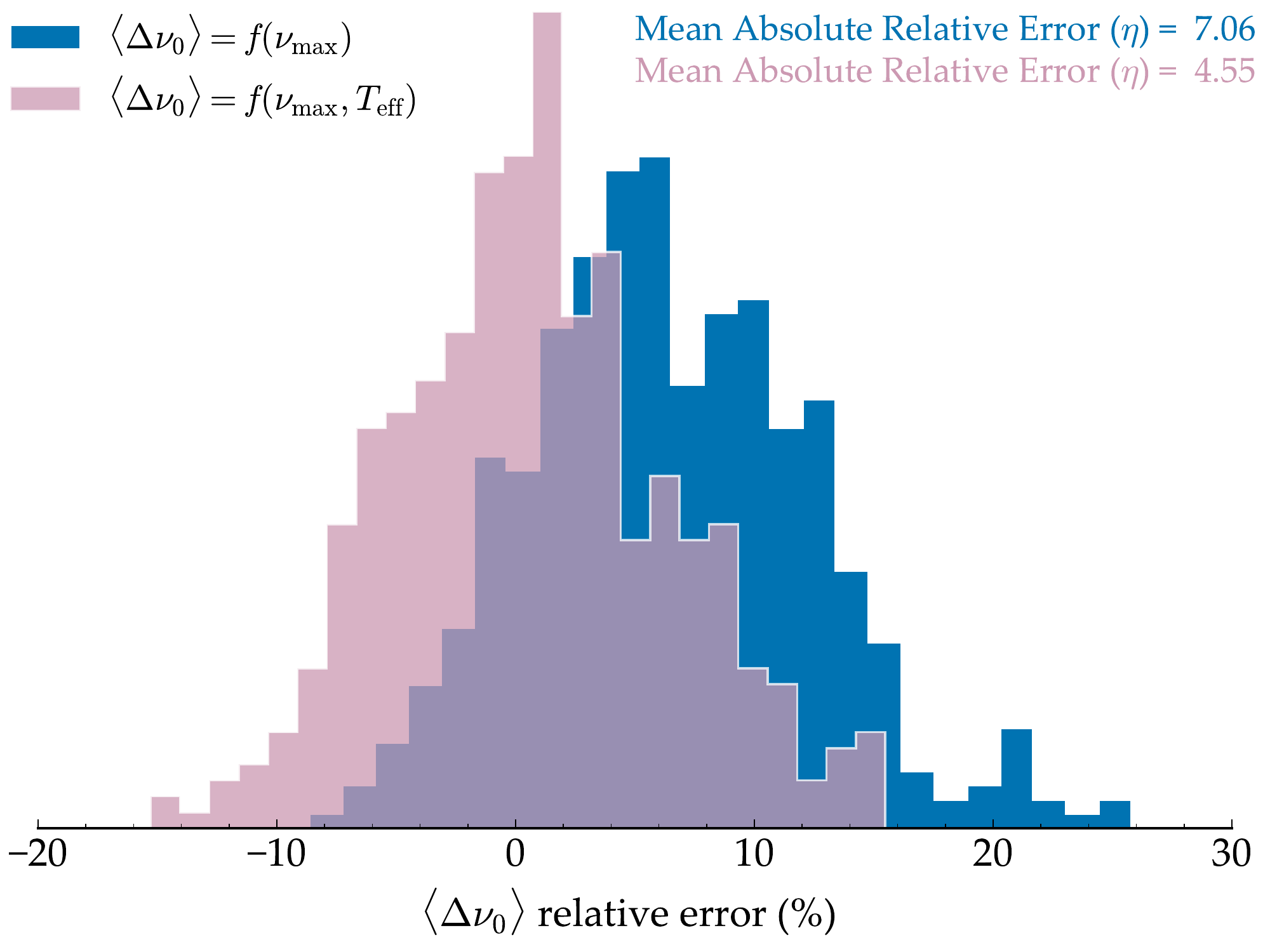}
\caption{Relative error (\%) in the predictions for $\langle\Delta\nu_0\rangle$ for 467 stars reported in \citet{2014ApJS..210....1C}. The blue colored distribution indicates the error in the predictions from the random forest using $\nu_{\rm{max}}$ as the only input observation whilst the distribution marked in lavender are the results from providing $\nu_{\rm{max}}$ and $T_{\rm{eff}}$. In the calculations we employ the effective temperatures determined from \citet{2012ApJS..199...30P} based on  SDSS photometry.} 
\label{fig:chap}
\end{figure}

In lieu of a direct measurement, $\langle\Delta\nu_0\rangle$ can be estimated from stellar models via an asteroseismic scaling relation (Equation \ref{equ:dnu}). Alternatively, it may be inferred from the observables through an empirical power law that relates $\langle\Delta\nu_0\rangle$ to $\nu_{\rm{max}}$ \citep{2009A&A...506..465H,2009MNRAS.400L..80S}. 
The power law estimates $\langle\Delta\nu_0\rangle$ within  15\,\% of its measured value \citep{2009MNRAS.400L..80S}.
We compare the RF's ability to likewise predict $\langle\Delta\nu_0\rangle$ from $\nu_{\rm{max}}$ in Table \ref{tab:dnu}. We also consider two and three parameter combinations for inferring $\langle\Delta\nu_0\rangle$ with the requirement that they do not comprise the remaining seismic observables. 

We find that the RF predicts $\langle\Delta\nu_0\rangle$  from $\nu_{\rm{max}}$  with $\mu (\eta) \approx 6\,\%$. These results are based on error free information (cross-validation hence no measurement noise) and the inclusion of $\nu_{\rm{max}}$ from a scaling law. 
In order to conduct a more faithful comparison with \citet{2009MNRAS.400L..80S}, we analyse the same data used in the derivation of their power law.  
Their Table 1 is a compilation of  $\nu_{\rm{max}}$ and $\langle\Delta\nu_0\rangle$ values from the literature. The data are predominately from radial velocity studies and measured with less precision than we have come to expect from \emph{Kepler} timeseries; they provide a robust test of the RF. 
We feed the RF the quoted $\nu_{\rm{max}}$ measurements and predict associated $\langle\Delta\nu_0\rangle$ values. We compare our predictions to the $\langle\Delta\nu_0\rangle$ values from the literature which are used to calculate corresponding 
$\epsilon$ and $\eta$ scores. Our results are presented in Table \ref{tab:stello}. 
We omit entries from  the \citet{2009MNRAS.400L..80S} dataset that are outside the parameter ranges of our training grid.
For the remaining 17 stars we find $\mu (\eta) \approx 8\,\%$ which is comparable to  $\mu (\eta) \approx 6\,\%$ accuracy achieved from cross-validation test (approximately 15\,000 stars).

The last column in Table \ref{tab:stello} indicates that the accuracy from the RF is similar to that of the power law. In addition, we find that parameterizing the RF regression as a function of two observables reduces the uncertainty by a factor of 2-3   (Table \ref{tab:dnu}).
 This hints that the inclusion of a temperature or metallicity dependence may also improve the fit offered by the power law\footnote{Symbolic regression will help determine whether, in this case, the fitting by the RF has a sensible functional form that can be straightforwardly expressed by two independent variables. This result seems reasonable as the additional information is likely providing a better handle on the stellar mass.}. 
 
 Analysis of recent \emph{Kepler} data yields a similar result. In Figure \ref{fig:chap} we present the \added{percentage} error in our predictions of 467 stars measured by \emph{Kepler} as reported in Table 1 of \citet{2014ApJS..210....1C}. We analyse stars for which $\nu_{\rm{max}}$, $\langle\Delta\nu_0\rangle$ have been measured from the oscillation spectra along with  $T_{\rm{eff}}$ as determined by \citet{2012ApJS..199...30P} based on Sloan Digital Sky Survey (SDSS) photometry. Results from the \emph{Kepler} sample confirm that predictions for $\langle\Delta\nu_0\rangle$ are improved with the inclusion of $T_{\rm{eff}}$ (lavender distribution). The blue distribution indicates that $\langle\Delta\nu_0\rangle$ is systematically overestimated when the RF only has access to information from $\nu_{\rm{max}}$ -- a bias that may very well be present in the power-law fit. With the inclusion of    
  $T_{\rm{eff}}$ our predictions become more accurate and precise with the bias from the single parameter function mitigated. We do not quite reproduce the accuracy achieved in the cross validation (Table \ref{tab:dnu}) using error free information.  Unsurprisingly,  measurement uncertainty, which we do not consider here, does not permit the accuracy attained in the ideal case.

\subsubsection{The frequency of maximum oscillation power -- $\nu_{\rm{max}}$} 

\begin{table}[h]
\centering
\caption{Combinations of observables that  best constrain $\nu_{\rm{max}}$.}
    \begin{tabular}{ccccc}
\hline    \hline
\multicolumn{2}{c}{Parameters} && $V_e$ &$\mu (\epsilon)$ [$\mu$Hz] \\ \hline 
$\langle\Delta\nu_0\rangle$ &&&0.923  & 7.88 \\
$\log \rm{g}$ & $[\rm{Fe/H}]$ && 0.888 & 9.99 \\
$\log \rm{g}$     & $\langle\Delta\nu_0\rangle$     &&0.954  & 5.38\\
$T_{\rm{eff}}$     & $\langle r_{10}\rangle$ &   &0.960  & 5.11\\
$[\rm{Fe/H}]$    &$\langle\Delta\nu_0\rangle$        &       &0.992  & 2.90\\
$T_{\rm{eff}}$     & $\langle\Delta\nu_0\rangle$    &  &0.992  & 2.84\\
$\log \rm{g}$  & $T_{\rm{eff}}$                 &&0.999  & 0.83\\
\hline
    \end{tabular}
    \label{tab:nmx}
\end{table}

Currently we are unable to predict the frequency of maximum oscillation power from first principles. \citet{1991ApJ...368..599B} and \citet{1995AA...293...87K} showed that this quantity does scale with the acoustic cut-off frequency and can thus be estimated via the Equation \ref{equ:nmax} scaling relation. It is therefore expected that Table \ref{tab:nmx} indicates that  $\nu_{\rm{max}}$ is best inferred from $\log \rm{g}$  and $T_{\rm{eff}}$. These are the two observables that correlate strongest those parameters used to calculate $\nu_{\rm{max}}$ in the training grid.

\subsubsection{The small frequency separation -- $\langle\delta\nu_{02}\rangle$}
The small frequency separation is an indispensable piece of independent information for determining stellar age. In the asymptotic limit \citep{1980ApJS...43..469T}
\begin{equation}
\langle\delta\nu_{13}\rangle = \frac{5}{3} \langle\delta\nu_{02}\rangle 
\end{equation}
and as Table \ref{tab:d02} demonstrates, the RF recovers  $\langle\delta\nu_{02}\rangle$ in the unlikely case that it is not extracted but $\langle\delta\nu_{13}\rangle$ is.
If we disregard combinations that include the seismic ratios, which also contain information of the local small frequency separation, we lack sufficient information to satisfactorily constrain $\langle\delta\nu_{02}\rangle$. 
Clearly much of the evolutionary aspect of this quantity can be explained though parameters 
that correlate with main-sequence lifetime e.g.,  $\log \rm{g}$, $\langle\Delta\nu_0\rangle$,   $\nu_{\rm{max}}$ and  $T_{\rm{eff}}$. However the associated errors of $\mu (\epsilon) > 1.0 \ \mu$Hz can correspond to large age uncertainties for main sequence stars ($\eta > 10\,\%$).

\begin{table}[h]
\centering
\caption{Combinations of observables, without the asteroseismic rations, that best constrain $\langle\delta\nu_{02}\rangle$.}
    \begin{tabular}{cccccc}
    \hline
\multicolumn{3}{c}{Parameters} && $V_e$ & $\mu (\epsilon)$ [$\mu$Hz] \\ \hline \hline
$\langle\delta\nu_{13}\rangle$ &  &   &&0.944 & 0.66 \\ 
$\log \rm{g}$ &  &   &&0.542 & 2.08 \\ 
$\langle\delta\nu_{13}\rangle$ &  $\langle r_{10}\rangle$  &   &&0.987 & 0.320 \\ 
$T_{\rm{eff}}$     & $\nu_{\rm{max}}$&     &&0.776  & 1.40\\
$\log \rm{g}$     &  $T_{\rm{eff}}$&                 & &0.775  & 1.40\\
$\log \rm{g}$     &$\nu_{\rm{max}}$ &               &&0.772  & 1.41\\
$\log \rm{g}$     &$\langle\Delta\nu_0\rangle$ &               &&0.723  & 1.54\\
$T_{\rm{eff}}$     & $\langle\Delta\nu_0\rangle$&     &&0.720  & 1.58\\
$\log \rm{g}$ & $[\rm{Fe/H}]$ &  && 0.720 & 1.59 \\
$\log \rm{g}$ & $[\rm{Fe/H}]$ & $\langle\Delta\nu_0\rangle$ && 0.861 & 1.06 \\
$\log \rm{g}$ & $\nu_{\rm{max}}$ & $\langle\Delta\nu_0\rangle$ && 0.860 & 1.09 \\
\hline
    \end{tabular}
    \label{tab:d02}
\end{table}

\section{Quantifying the Required Measurement Accuracy of Stellar Observables}
\label{sec:accu}
In the previous section we used RF regression to appraise how well combinations of observables constrain other stellar parameters. The $\approx 50\,000$ RFs were   
evaluated using cross-validation. The tests are a pure measure of the regressor's performance as we have error-free information that we attempt to reproduce (withheld models).
 As we have already alluded to, like all procedures that seek to infer stellar parameters, we must also consider the consequences of measurement uncertainty in our method.

Measurement uncertainty will impact the RF results in a manner that is different to model finding algorithms. Consider an iterative model finding procedure in which we seek an  optimum model for a set of observations. We can typically expect $T_{\rm{eff}}$ as a constraint with an associated  uncertainty of $\sigma =100\,K$.  The minimization algorithm will identify a set of candidate models, many with quite different structures. Hence the uncertainty in $T_{\rm{eff}}$  will impact 
all stellar quantities simultaneously.  The  RF, on the other hand,
 builds a statistical description of stellar evolution by calculating a regression model for each individual parameter from the training data. 
 The BA1 method requires that each input observable is perturbed with random Gaussian noise according to its measurement uncertainty. Monte Carlo perturbations are performed 10\,000 times and each instantiation evaluated by the RF to yield individual density distributions for each stellar parameter. Thus the uncertainty in $T_{\rm{eff}}$, or any observable for that matter, will only impact on the predictions 
 of each parameter in proportion to the degree to which it features in that parameter's regression model.

 The methodology, combined with the speed of the RF, provides a tractable means to 
 asses how the individual measurement uncertainty of an observable will impact upon each predicted stellar quantity. We hence determine  how accurately the observables must be measured in order to achieve a desired precision from the RF.

\begin{figure*}
    \centering
    \includegraphics[width=0.95\textheight,height=0.95\textwidth,keepaspectratio, angle=90]{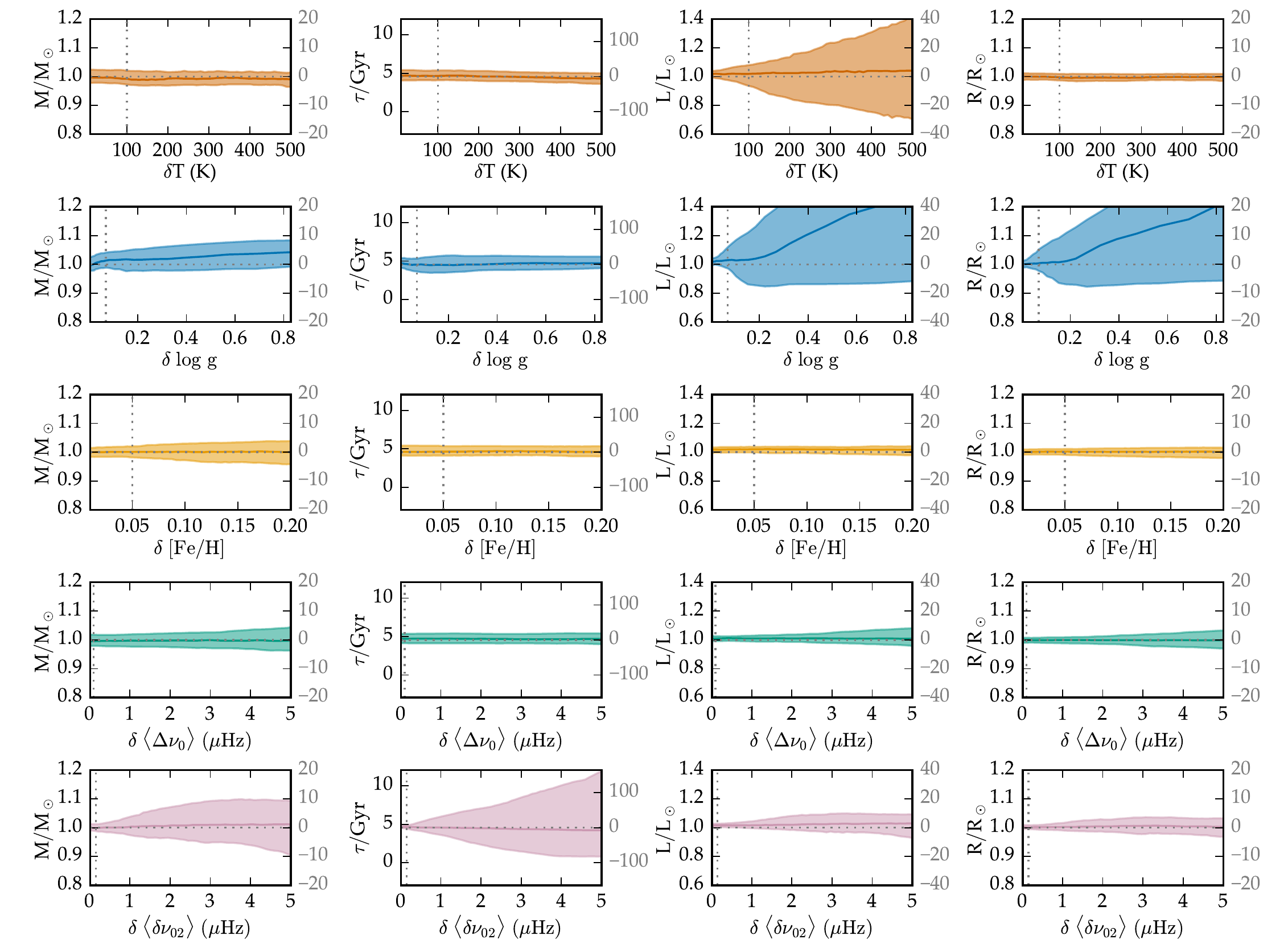}
    \caption{Predictions for the solar mass, age, luminosity and radius as a function of the uncertainties applied to key observables. \added{In each panel we have perturbed the quantity on the abscissa in isolation, centred around the measured value listed in  Table \ref{tab:sun} and with the uncertainties in the ranges specified therein. We indicate the median predicted value (solid line) and the 68\%  confidence interval (shaded region).} The dotted horizontal lines mark the zero point or true value in each panel and the vertical line indicates a typical observational uncertainty for the perturbed quantity.}
    \label{fig:uncert1}
\end{figure*}

We train a RF on the observables listed in Table \ref{tab:sun}.
We take the (approximate) solar value of each observable as our measurement and consider
`observational uncertainties' ($\sigma$) within the ranges specified in Table \ref{tab:sun}. 
We first perturb the measurement values with Gaussian noise assuming the minimum $\sigma$ values listed. We produce 10\,000 instantiations for that set of $\sigma$ values, ensuring each perturbed observable remains within the limits of our training grid.
We evaluate stellar parameters and determine detailed distributions for that set of uncertainties. We repeat the process increasing the $\sigma$ for a single observable  always keeping the  $\sigma$ values of the other observables at their minimum. 
We draw 50 $\sigma$ values for each observable sampling their specified ranges evenly.  We produce  probability density distributions for 250 sets of $\sigma$ values, the results of which are summarised in Figure \ref{fig:uncert1}.

 \begin{table}
    \caption{Central values and uncertainty ranges used for predicting the Sun in Figure \ref{fig:uncert1}.}
    \begin{tabular}{lccc}
    \hline \hline
Quantity & Value & Min($\sigma$) & Max($\sigma$) \\ \hline 
$T_{\rm{eff}}$ (K)  & 5777 & 10  & 500\\
$\log \rm{g}$ &  4.43812 & 0.00013 & 1.0\\
$[\rm{Fe/H}]$ & 0.0 & 0.05 & 0.2 \\
$\langle\Delta\nu_0\rangle$ ($\mu$Hz) & 136.0 & 0.5 & 10\\
$\langle\delta\nu_{02}\rangle$ ($\mu$Hz)& 9.0 & 0.5 & 5 \\
\hline
    \end{tabular}
    \label{tab:sun}
\end{table}

In Figure \ref{fig:uncert1} we plot the median value (solid line) and the 68\%  confidence interval (shaded region)
 for $M$, $\tau$, $L$ and $R$ as a function of the uncertainty applied to each observable. The figure is organised such that each row (and color) corresponds to the observable that has had its uncertainty increased and each column corresponds to the model parameter of interest. In this Figure, the left axis indicates the predicted value from the RF and the right axis indicates the relative error with reference to the true values of the Sun.  The horizontal dotted grey lines mark the reference value in each case whilst the dotted vertical lines indicate a typical uncertainty for the perturbed observable. 

The particular RF we have trained does not significantly rely on $T_{\rm{eff}}$ in its regression model for $M$, $\tau$ or $R$. As the radius is supplemented by the seismic quantities, any uncertainty in $T_{\rm{eff}}$ is propagated as uncertainty in the luminosity. We find a typical uncertainty of 100\,K corresponds to an error of $\pm 0.2$ $L$/L$_{\odot}$ at the 68\% confidence level.

The inference on solar mass is affected once $\delta \, \log \rm{g} > 0.03$. 
However, even at unreasonably large values of  $\delta \, \log \rm{g} = 1$, the uncertainties for mass and age remained relatively constrained by additional seismic information. We find 
that $L$ and $R$ are far more reliant on $\log \rm{g}$ in their regression function with uncertainties in these quantities growing significantly once  $\delta \, \log \rm{g} > 0.1$.

The feature importances in BA1 indicate that $[\rm{Fe/H}]$ is used most often by the RF in  crafting its decision rules.  The four stellar parameters we investigate here indeed all rely on information from $[\rm{Fe/H}]$, however, they are supplemented by seismic information which helps to constrain the uncertainty in their predictions. It is the model parameters such as the mixing length, degree of overshoot and initial metallicty  that become much less certain as we increase $\sigma([\rm{Fe/H}])$ (not shown here).

The seismic diagnostics are very sensitive to the stellar structure, and hence also those  parameters we use to characterize a star ($M$, $\tau$, $L$ and $R$). We have seen how reliant the RF is on the seismic diagnostics in the  regression models, allowing us to still predict the structural properties with relatively good precision in the face of large spectroscopic uncertainties. Without accurate measurement of $\langle\Delta\nu_0\rangle$ the uncertainty in structure parameters increase significantly. Whilst the uncertainty in $\langle\delta\nu_{02}\rangle$ does introduce some small uncertainty in $M$, $L$ and $R$, as expected, its accuracy significantly impacts upon our ability constrain stellar age.

\section{Discussion}
\label{sec:disc}

Advances in stellar evolution theory are usually sought through refinement of the standard canonical model.  In this classical approach, observations reveal  behaviour that cannot be explained by the current stellar theory, a model is constructed, analysis of the resultant predictions are carried out and conclusions on the efficacy of that model drawn. In this study we adopted a complementary approach: an exploratory based method whereby we performed statistical analysis of models covering a large range of known physics. Rather than first develop a new model to evaluate, we explored the current paradigm to quantify existing relationships and draw new conclusions.  

Some of the techniques employed in this analysis are over 100 years old and in many areas of research are powerful standalone tools. They have rarely featured in the field of stellar modelling. Here we comment briefly on the timing of our manuscript which we attribute to two main factors: the advent of supervised machine learning techniques and modern computing resources.  

Random forests are an integral part of the present analysis and are a modern technology.
They help place the use of statistical methods in stellar evolution in a wider practical context.  Elucidating both the relationships found by RF and the exploitable information inherent in the model data provided motivation for the use of techniques such as PCA and correlation analysis.  The RF further facilitated the application of these methods due to the requirement that the models be cast  into a comprehensive evolutionary matrix; something that is not strictly necessary for grid based searches.     

Our approach shares similarities to that taken by \citet{Brown1994} although we differ in methodology. Since their work, we have seen the necessary increase in computing power and the success of the \emph{Kepler} and CoRoT space missions.  The statistical analysis here requires a well sampled grid of stellar models both with structure and oscillations computed. It cost a week of modern supercomputing time to generate the matrix upon which these operations are performed. Evaluating and training approximately 50\,000  RFs itself is also a computationally expensive endeavour.

\subsection{Features of the Dataset}
It is not clear \emph{a priori} through inspection of the equations of stellar structure, if and how any two emergent quantities of the models co-vary. 
There are, of course, combinations of parameters whose covariances are well-founded in stellar theory, but there exist quantities whose diagnostic power remain underutilized and could in fact offer additional insight into the underlying models.
Bringing such relationships to light over the collective lower main sequence is a key aim of our statistical investigation. 
The correlations in the truncated grid (Figure \ref{fig:filt_corr}) and full BA1 grid (\ref{fig:corr}) reveal the relationships that can be utilized to constrain each of the quantities listed in Table \ref{tab:parmdefs}.
Many of the model properties that we wish to infer correlate with several observables simultaneously.  This indicates that the observables carry redundant information about the star. 
In addition, observables co-vary amongst themselves. During  iterative model searches some of the covariances, such as between the seismic ratios, are taken into account.  However, for example, it is possible to obtain independent measurements of $\nu_{\rm{max}}$, $\langle\Delta\nu_0\rangle$, and $\log \rm{g}$. Treating these as independent degrees of freedom without considering model covariances then biases the fit towards the parameters to which these quantities pertain and can result in a solution that is overfit. 

We determined the degree of degeneracy in the observables through PCA dimensionality reduction. 
As mentioned previously, RF regression falls under the umbrella of \emph{supervised} learning, whereas PCA is a form of \emph{unsupervised} learning. 
The difference is that in supervised learning, there is a correct answer that the algorithm is trying to understand how to reproduce.
In the case of unsupervised learning, the machine attempts to directly infer properties of data without any help from the supervisor. Hence, regression and classification analyses are forms of supervised learning, whereas cluster and factor analyses are examples of unsupervised learning.
In the case of supervised learning there is a clear measure of success in the resultant model. 
There is a desired output that the inputs try to match. 
The efficacy can be quantified and evaluated via, say, cross-validation or information-theoretic metrics. 
Unsupervised learning methods simply try to identify features and in the case of PCA these features are not necessarily interpretable. 

The PCA in \S \ref{sec:PCA} focused on the truncated grid. It comprises 11 stellar observables of all which carry information on the model properties to varying degrees. 
 We found that 99.2\,\% of the variance in the observables could be explained by five components with
nearly 98\,\% of the data are explained by four components. It could be argued that PC$_5$ explains noise rather than features, however, we found that PC$_5$ displays distinct enough correlations (i.e., with near surface physics) that it warrants inclusion in our analysis. 
The clear dimensionality reduction, from 11 observables to five PCs,  highlights the value in performing PCA: had we found comparable contributions from each component, we would have instead confirmed a clear dominance from higher order relations and an inadequacy of an approach based on linear analysis.

Our primary goal in \S \ref{sec:PCA} was to reduce the dimensionality of the observables.
We initially considered regions of the parameter space where observations have shown stars to occupy. 
Following on from the rank correlation tests in \S \ref{sec:RCT} we applied PCA to a truncated version of the BA1 grid.  
However, the results of the PCA depend on the properties of the data and will change depending on features such as the parameter ranges and number of models in the grid. 
For example performing PCA on the full set of evolutionary tracks (340800 models) demands that components are dedicated to explaining variance in (wider) unobserved regions of the parameter space. 
In order to demonstrate that our interpretations of the PCs are robust, we repeated the PCA on four different subsets of the BA1 grid. 
 We made cuts to the mass and metallicity ranges on the training data the results of which are included in Appendix \ref{sec:PCAg} by means of qualitative correlation plots. 
 
The PCs of the respective grids explain a similar percentage of the variance in each grid: PC$_1$ accounts for approximately 40\% of the variance,  PC$_2$  approximately 35\%  etc.,  with more than 75\% of the variance in the observables explained by the first two PCs. 
We interpret this result as the PCA capturing essentially the same five inherent `features' in the observables. 
It follows that the choices in grid size and parameter range have only a small effect on the explained variances.
Analysis of all four grids helps further illustrate that there is redundant information carried in some observables, particularly the seismic separations and ratios. 
Varying the parameter ranges changes the correlations between the PCs and observables (loadings) yet the PCs still explain a similar percentage of the variance in each case.  
Due to the information redundancies the PCs can be constructed such that same features are captured with different linear combinations of the observables.
How exactly a PC is constructed in a particular grid will depend on the amount of variance in the observables imparted by the chosen parameter ranges.

With respect to the independent model parameters, it is no surprise that in general PC$_1$ is strongly correlated with the stellar mass ($M$) and  and PC$_2$ with initial metallicity (Z$_0$). 
These are the principle determinants of stellar evolution in that order and both impact upon the stellar structure independently.
In the two grids where we have cut the mass and metallicity ranges we find that the loading of $T_{\rm{eff}}$ is larger in PC$_1$.
This is because in the more solar-like tracks $T_{\rm{eff}}$ is a strongly monotonic function of evolution.
The surface aspect of PC$_2$ is then supplemented with some information from  $\log \rm{g}$ and $[\rm{Fe/H}]$.

Reducing the dimensionality of the observables and relating them back to the model parameters without redundancy aided with the interpretation of the PCs. 
Whilst it is useful to have the observables so succinctly described, it does not provide insight into the model parameters we wish to infer. 
We thus condensed the information from the correlation plots into a $\Lambda$ score which 
is the sum of the square of the correlation coefficients  between the model parameters and the PCs (determined for the observables). Squaring the correlation coefficients is equivalent to the squaring the PC loadings of the centered and scaled observables. 
The score is a means to quantify the extent to which information from the model parameters, dependent and independent, are encoded in the observables.
We calculated $\Lambda$ scores for all four grids upon which PCA was performed (Appendix \ref{sec:lambdaa}) and indeed found mostly consistent results. We note some differences arise in the initial model parameters such as $\alpha_{\text{MLT}}$ and $\alpha_{\text{OV}}$ which reflect their underlying 
distributions from the choices in grid truncation. The above analyses can be applied to any combination of observables and model parameters to gauge their utility.

\subsection{Exploiting the Inherent Relationships}

Understanding the inherent properties of the collective lower main sequence is the first step in elucidating the BA1 RF regression. The statistical analysis quantified what information was present in the training data for the RF to exploit. We illustrated why the available data permit BA1 to predict parameters such as $M$, $R$ and $L$ with such high precision and why initial model parameters such as $D$ and $\alpha_{\text{MLT}}$ remain uncertain in comparison. Whilst  \S \ref{sec:RCT} and \S \ref{sec:PCA} demonstrated the breadth of information available to the RF, in \S \ref{sec:qu} we determined how the information could best be used. 

RFs are amongst the most powerful tools in mathematics for non-linear regression. 
The BA1 RF uses the observables, creating a set of decision rules that reduce the variance in the parameter it is fitting.
Whilst feature importances provide some insight into this process as a whole it does not provide specific details for the individual parameters. 
By performing non-parametric multiple regression with every combination of observable in our grid,  
we demonstrated how the correlations in Figure \ref{fig:filt_corr} could best be exploited and best combined to reveal the most information about each stellar quantity.
Two of the observables, $[\rm{Fe/H}]$ and $\langle\delta\nu_{02}\rangle$ (or as a ratio),  are of vital importance in model fitting procedures as they provide indispensable pieces of independent information that cannot be inferred from other quantities.  

We in effect invert the observations for the model parameters based on functions learnt from the training data. 
Thus we can determine the relative importance of each observable for inferring the model parameters. We, in addition, provide a precision with which we can determine each model parameter \emph{directly} from the information contained in the observables. 
The attainable precision is a function of the number of initial model parameters that are varied and the model degeneracy in the data. 
For example, with perfect information from the observables, the six dimensions in the BA1 grid limits our inference on mass to 
$\mu (\epsilon) =0.02 \ \Mo$.

Many of the Tables in \S \ref{sec:qu} demonstrated an important property of the RF. 
In the case of missing or unreliable measurements of an observable, the RF can draw upon information redundancies in the data to determine new regression rules for the model parameters.
In principle, such redundancies can lead to biases and overfitting in iterative model finding methods.
During such search procedures the best fitting stellar model is the one that best matches all of the observations but each observation only bares on some parts of the model, and observations can contain redundant information.

Through statistical bagging and multiple regression the RF avoids the problem of overfitting altogether. 
These underlying methodologies are the reason why in \S \ref{sec:accu} many of the parameters we inferred remained well constrained despite large uncertainties in some of the observables.   
In statistical bagging different subsets of the training grid are sent to different nodes. Each node will use information theory to create a set of decision trees to explain the parameter of interest. The nodes will differ in their rules and choice of parameters. Thus the uncertainty in an observable will only impact on the parameter we infer to the extent to which the observable is used in the rules.  
Take the example from Figure \ref{fig:uncert1} where with a 5 $\mu$Hz uncertainty in $\langle\Delta\nu_0\rangle$ the RF still predicts the solar properties albeit with slightly less confidence.
The other observables help constrain the predictions. 

Part of the analysis in \S \ref{sec:qu} demonstrated the best possible (average) precision in which we can hope to infer stellar parameters. Our error analysis in \S \ref{sec:accu} is an extension of this. Rather than assume perfect information we determined  
 the measurement accuracy required of the observables to attain a desired precision from the RF.
Our analysis focused on the Sun and is indicative of solar-like analogues.
In Table \ref{tab:relabunds} we saw some of the large uncertainties associated with retrodicting abundances in low-metallicity stars. We have greater degeneracy with the efficiency of diffusion and the initial abundances. These large error scores by no means indicate that the RF is incapable of characterizing low-metallicity stars. Rather it is an honest appraisal of stellar uncertainties when we do not make assumptions of the initial abundance say through a dY/dZ  chemical evolution ``law"  or a fixed diffusion efficiency. Our error analysis here does not take into account covariances and was designed to investigate the impact on a observable-by-observable basis.
A more detailed error analysis and the associated issues at low metallicity form the focus of a forthcoming paper.

\subsection{Implications for the \emph{TESS} and \emph{PLATO} missions}
\begin{figure*}
    \centering
    \includegraphics[width=\textwidth,height= \textheight, keepaspectratio ]{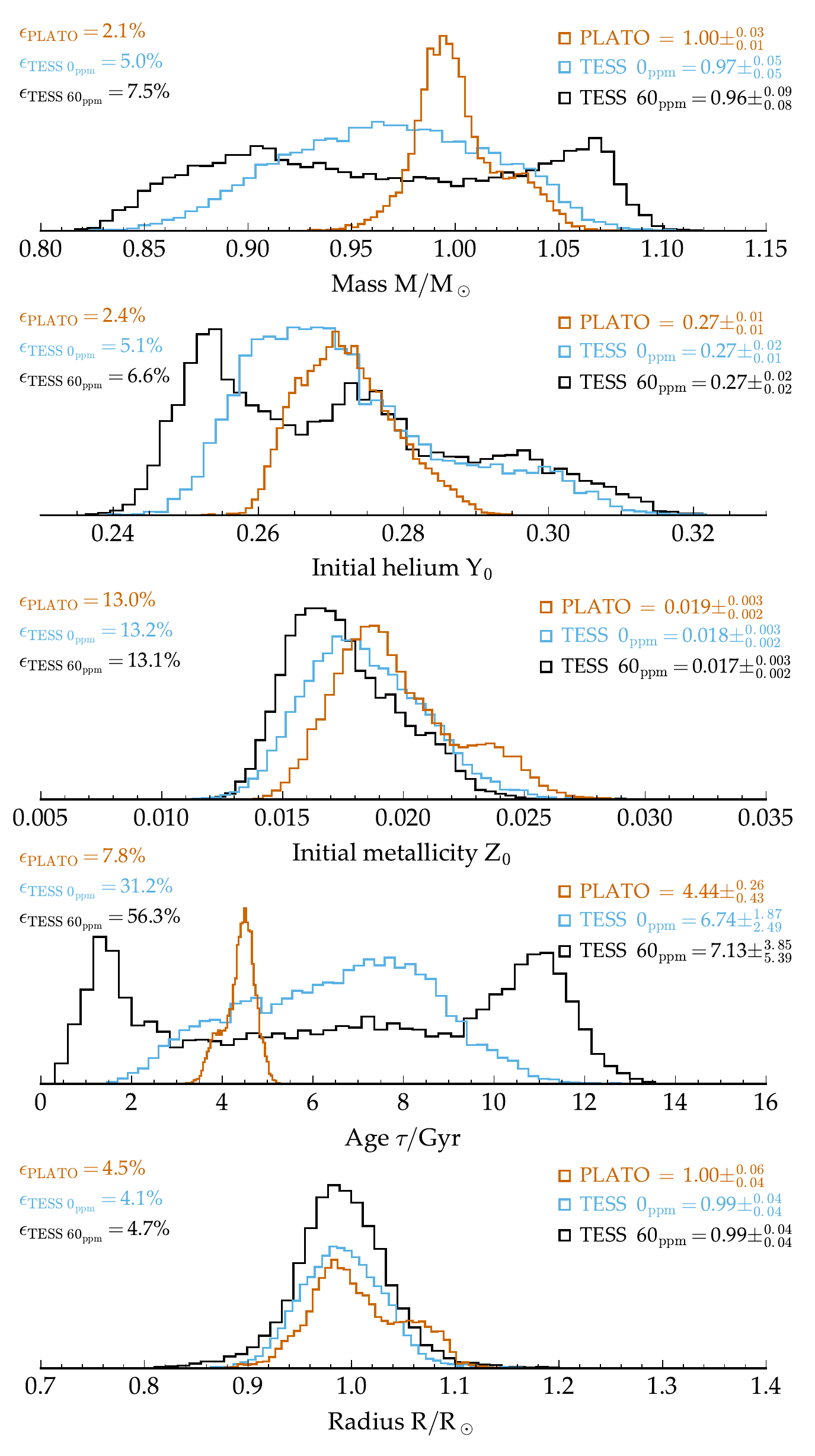}
    \caption{Predictions for the `Sun as a star' using observations expected for targets from \emph{TESS} (assuming two different systematic noise levels) and \emph{PLATO} space missions. In each panel we list the median with uncertainties (84\%-50\% confidence intervals and 50\%-16\% confidence intervals)   for the quantities as well as the relative error in our prediction. 
    } 
    \label{fig:TessPlato}
\end{figure*}

\begin{deluxetable*}{lccccccccc}
\tablecaption{Solar data degraded to the level expected for sun-like stars in: the \emph{TESS} catalogue assuming systematic  noise of 60 ppm hr$^{1/2}$ from the mission, \emph{TESS} assuming no systematic noise and from \emph{PLATO}. For each set of observables we include the feature importances from the random forest used in characterising the 'Sun as a star'. Note that in the case of the expected \emph{PLATO} data we have perturbed a subset of frequencies according to their distance from $\nu_{\rm{max}}$. The numbers reported for the separations and ratios are thus the respective means and standard deviations of 10,000 perturbations to the data which we evaluate to determine our parameter distributions. \label{tab:tpl}} 
\tablehead{
        \multicolumn{1}{c}{} &
        \multicolumn{3}{c}{TESS (60 ppm hr$^{1/2}$)}   &
        \multicolumn{3}{c}{TESS (0 ppm hr$^{1/2}$)}   &
        \multicolumn{3}{c}{PLATO} \\
        \colhead{Parameter}  &
        \colhead{Value} &
        \colhead{Uncertainty} &  
        \colhead{Importance} &
        \colhead{Value} &
        \colhead{Uncertainty} &
        \colhead{Importance} &
        \colhead{Value} &
        \colhead{Uncertainty} &
        \colhead{Importance}
}
\startdata
$T_{\rm{eff}}$ (K) & 5778 &100 &29.3\% & 5778 &100 &26.7\%& 5778 &100 &16.2\%\\
$[\rm{Fe/H}]$ &-0.014 &0.021 &34.3\% &-0.014 &0.021 &33.4\% &-0.014 &0.021 &27.9\%\\
$\log \rm{g}$ &4.43 &0.07 &18.5\% &4.43 &0.07 &12.4\% &4.43 &0.07 &8.8\\
$L \ (L/\rm{L}_{\odot})$   &0.98  &0.04 &18.0\% &0.98  &0.04  &16.7\% &0.98  &0.04  &7.8\%\\
$\nu_{\rm{max}} \ (\mu \rm{Hz})$  & -- & -- & -- &3093 &100 &10.8\% & -- &-- & --\\
$\langle\Delta\nu_0\rangle  \ (\mu \rm{Hz})$  & -- & -- & -- & -- & -- & -- &134.81 &0.05 &6.4\%\\
$\langle\delta\nu_{02}\rangle  \ (\mu \rm{Hz})$  & -- & -- & -- & -- & -- & -- &9.02 &0.15 &7.1\% \\
$\langle r_{01}\rangle$ & -- & -- & -- & -- & -- & -- &0.0226 &0.0005 &7.4\%\\
$\langle r_{10}\rangle$ & -- & -- & -- & -- & -- & -- &0.0227 &0.0005 &7.3\% \\
$\langle r_{02}\rangle$& -- & -- & -- & -- & -- & -- &0.0668 &0.0011 & 11.1\%
\enddata
\end{deluxetable*}

The NASA \emph{TESS} mission \citep{2015JATIS...1a4003R} and ESA's \emph{PLATO} \citep{2014ExA....38..249R} herald a new age for the space-based photometry and the detection of planetary transits. Due to launch in \added{2018} and 2025 respectively, their common primary science mission is to identify terrestrial planets around bright stars. The pre-selection of bright targets will ensure that the stellar hosts can be further analyzed with spectroscopy and it is expected that many of the planet candidates will be suitable for atmospheric follow-up (ideally) with the James Webb Space Telescope. As was the case with the \emph{Kepler} and \emph{CoRoT} missions, the photometric time-series observations will prove useful to asteroseismology. In the case of \emph{PLATO} the study of the stellar structure through asteroseismology is a key science goal in the mission design \citep{2014ExA....38..249R}.  

\emph{TESS} will monitor photometric variations of $> 10^5$ low-mass main-sequence stars. Under its `step and stare' pointing strategy, fields will be monitored for periods ranging from one month to one year depending primarily on their ecliptic latitude. With its two minute and 30 minute cadences, \emph{TESS} will be able to detect small rocky planets around solar like stars at $\le$ 7th magnitude. It is expected to detect of the order 1700 planets with sub-Neptune masses \citep{2016ApJ...830..138C} and will identify many more larger planets around dimmer targets.  The asteroseismic potential of TESS has been rigorously investigated by \citet{2016ApJ...830..138C}. Their analysis of the expected \emph{TESS} photometry indicates the presence of an oscillation power excess 
in low-mass main-sequence stars when there is no systematic noise present in the data. With an expected systematic noise level of  60 ppm hr$^{1/2}$ from the mission, their analysis indicates a detectable power-excess in F-dwarfs as well as sub giants and red giants -- this owing to the higher luminosity and hence larger mode amplitudes in these stars. For a majority of stars the 27 day pointing is insufficient to extract detailed asteroseismic diagnostics such as mode frequencies or separations. Rather, the seismic information will be limited to the determination of $\nu_{\rm{max}}$ in stars where the power-excess is detected. As a consequence, masses and radii for the \emph{TESS} targets are to be determined using a combination of  GAIA data, the  $\nu_{\rm{max}}$ -- $\langle\Delta\nu_0\rangle$ power law \citep{2009A&A...506..465H,2009MNRAS.400L..80S} , asteroseismic scaling relations and grid-based searches.

The number of small planet detections from the \emph{PLATO} mission is expected to eclipse the number found by \emph{Kepler}
and \emph{TESS} by up to three orders of magnitude. In addition, the \emph{PLATO} pointing strategy will allow for the measurement of oscillation frequencies in $>$ 80,000 dwarf and subgiant stars with magnitudes less than 11.  In total the mission will provide approximately one million light curves for stars with brightness $\le$ 13th magnitude \citep{2014ExA....38..249R}. In many stars modes up to spherical degree $\ell =3$ will be detected with typical frequency uncertainties in the range  0.1 -- 0.3 $\mu$Hz. The second major science goal of \emph{PLATO} is to 
probe stellar structure and evolution by asteroseismology and provide support to exoplanet science through determining
\begin{itemize}
\item stellar masses with an accuracy of better than 10\%,
\item stellar radii accurate to 1--2\%, and
\item ages of solar-like stars accurate to 10\%
\end{itemize}

Here we treat the `Sun as a star' in order to quantify how well we can characterise target systems observed by the 
upcoming space missions and to determine the prospect of meeting the accuracy requirements.  
In Table \ref{tab:tpl} we indicate the observables the missions are likely to provide. 
We degrade the corresponding solar data according to the expected uncertainty from the respective measurements.   
 As \emph{GAIA} is complete down to 20th magnitude we have assumed that distances and hence luminosities will be available for all targets in these missions. We consider data for \emph{TESS} targets assuming both 60 ppm hr$^{1/2}$ and no systematic noise in the photometry. Thus in the case of the latter we anticipate that an oscillation power excess can be extracted for a solar-like star and $\nu_{\rm{max}}$ determined. The large and small frequency separations for the \emph{PLATO} data are determined by degrading a subset of solar frequencies using the method described in BA1. We take a conservative approach in this calculation and assume that the $\ell=3$ modes are not extracted.

Figure \ref{fig:TessPlato} shows our predictions for masses, radii, ages, initial helium and metallicity for a `Sun-as-a-star' exercise. In each panel we indicate the median of the probability density distribution and the corresponding  uncertainty from the 16\% and 84\% confidence intervals for the parameter we are predicting. In addition we determine the relative error which we define as $\epsilon = 100 \cdot \sigma/\mu$  where $\mu$ is the mean and $\sigma$ is the standard deviation of the distributions. In Appendix \ref{sec:PSM} we further demonstrate the impact of the measurement uncertainty on the prediction of each quantity as per Figure \ref{fig:uncert1}.

Although we can expect accurate mass determinations for targets in both missions, the supplementary seismic data from \emph{PLATO} allows us to improve the precision with which we determine mass by approximately a factor of two. This is despite the fact the RF has identified a less-likely but not impossible (slightly) younger, higher-mass, higher-metallicity solution from the  \emph{PLATO}  data (we find bimodalities for most quantities predicted with the \emph{PLATO} observables). In the case of \emph{TESS}, the absence of the large frequency separation leads to greater uncertainty. One of the methods discussed  by \citet{2016ApJ...830..138C} for the mass determination of \emph{TESS} targets is to use the power law linking $\nu_{\rm{max}}$ to $\langle\Delta\nu_0\rangle$ (which has been shown to be accurate to 10-15\%) and apply the asteroseismic scaling laws (Equations \ref{equ:nmax} and \ref{equ:dnu}).  In \S \ref{sec:seispr} we demonstrated that the random forest exploits further information from  temperature or metallicity measurements to improve the accuracy of the $\nu_{\rm{max}}$ -- $\langle\Delta\nu_0\rangle$ relation.  Thus we expect the accuracy with which we predict mass from \emph{TESS} data to represent an upper limit to that attainable by  applying the power-law and scaling relations. 

The assumption of \emph{GAIA} distances and hence stellar luminosities ensure that radii can be determined for targets in both missions -- the seismology is essentially redundant for the inference of the stellar radius.  We note that the relative error for  \emph{PLATO} in our `Sun-as-a-star' test  is a factor of two higher than the 1-2\% expected by the consortium. This is a consequence of having identified bimodal solutions. Their target accuracy can likely be met if the uncertainties in the measurements are further reduced and a unimodal solution found. 

The analysis in \S \ref{sec:seispr} has highlighted the necessity of the small frequency separation  in order to 
tightly constrain the ages of field stars. The predictions for age in Figure \ref{fig:TessPlato} are therefore as expected. 
The inclusion of oscillation frequencies and determination of the small frequency separation (and ratios) from \emph{PLATO} data result in age uncertainties for solar-like stars to within the 10\% level.  Without information from the core, ages for \emph{TESS} targets remain largely unconstrained and consistent with the accuracy typically expected when dating field stars spectroscopically.

\section{Conclusions}
In this work we examined the processes that allow random forest regression to rapidly and accurately infer stellar parameters \citep{2016ApJ...830...31B}. We shed light on the inherent properties of the model training data that the algorithm can exploit. 

\begin{itemize}
   
             \item We demonstrated that there is a large amount of information redundancy in the stellar parameters which is integral to the efficacy of the random forest algorithm. Through statistical bagging, the random forest  creates sets of decision rules using different combinations of observables to infer a given quantity. The methodology results in  robust predictions and includes the ability to compensate for data that are missing or unreliable. 
             
          \item We illustrated the behaviour of parameters across the collective lower main sequence with the relationships that arise (e.g., age -- luminosity) different to those that develop internally along an evolutionary track. This is the inherent information the random forest draws upon in its regression.
     
         \item  We found the parameter pairs that exhibit the strongest correlations correspond to well known asteroseismic and main-sequence relations.

   \item The random forest works well in cases when there is sufficient information and sufficient redundancy.
   Through principle component analysis we quantified the degree of degeneracy in the observables. 
   Our analysis demonstrated that 99.2\,\% of the variance in the 11 stellar observables could be explained by five principle components.

 \item The observables we have considered only carry five pieces of independent information.
             During  iterative model searches it is common that independently determined parameters such as $\nu_{\rm{max}}$, $\langle\Delta\nu_0\rangle$, and $\log \rm{g}$ are treated as independent degrees of freedom. The composition of the principle components indicate that by not considering their model covariances, any fit is biased towards the common stellar information to which these parameters pertain.

  \item We devised a score  which  allows us to rank the degree to which model parameters can be inferred from the observables.  Radius, luminosity, and main-sequence lifetime can be extracted  with confidence, however, the initial model parameters such as $\alpha_{\rm{MLT}}$, Y$_0$ and $\alpha_{\rm{ov}}$ are not sufficiently constrained by the observables and cannot be inferred directly form the data.  Our analysis can be extended in a straightforward manner to model parameters and observables not considered here.

  \item Having elucidated the statistical properties of the training data, we sought to better understand how the random forest uses the data in its decision making rules.
By performing non-parametric multiple regression with every combination of observable in our grid we determined:
\begin{enumerate}
    \item  which observables are the most important/useful for each model parameter,
    \item  the minimum set of observables that satisfactorily constrain each model parameter, and
    \item  the precision with which we can determine each model parameter \emph{directly} from the information contained in the observables. 
\end{enumerate}

\item We examined the quantities on a parameter by parameter basis and here highlight the results for mass and age.  In a grid of stellar evolution models varied in six initial parameters we find that the average error in predicting mass across the grid is $\pm 0.02 \ \Mo$ and  $\pm 282$ Myr for age. The average error in age increases by a factor of three when we are limited to information from only two observables such as in the Christensen-Dalsgaard diagram. Three parameters are sufficient for constraining mass whereas we require five observables to determine age.

\item We determined whether the random forest could reproduce the well-known power law that relates $\langle\Delta\nu_0\rangle$ to $\nu_{\rm{max}}$ and found that additional information from  $T_{\rm{eff}}$ or  $[\rm{Fe/H}]$ reduces the average error in the relation by a factor of two. 

\item We investigated the measurement accuracy required of the observables to attain a desired precision from the random forest. 
The processes of statistical bagging and multiple regression help mitigate the impact of large spectroscopic errors as the random draws upon complementary seismic information when devising its decision rules. 
The results confirm that $[\rm{Fe/H}]$  and  $\langle\delta\nu_{02}\rangle$ are indispensable independent pieces of information for model fitting algorithms.

\item Finally, we determined the accuracy and precision with which we can expect to characterise solar-like stars observed by the upcoming \emph{TESS} and \emph{PLATO} space missions. In both cases masses can be accurately inferred and measurements from \emph{GAIA} will ensure that radii are well constrained. Oscillation frequencies will not be detectable in most low-mass main sequence stars observed by \emph{TESS}. In contrast, the availability of the small frequency separation for \emph{PLATO} targets will permit accurately determined stellar ages. 
\end{itemize}

\acknowledgments 
The research leading to the presented results has received funding from the European Research Council under the European Community's Seventh Framework Programme (FP7/2007-2013) / ERC grant agreement no 338251 (StellarAges). E.B undertook this research in the context of the International Max Planck Research School for Solar System Research. S.B.\ acknowledges partial support from NSF grant AST-1514676 and NASA grant NNX13AE70G. We thank Alexy Mints and the anonymous referee for their useful comments and discussions which helped improve this manuscript.

\software 
Stellar models were calculated with  \emph{Modules for Experiments in Stellar Astrophysics} \textbf{r8118}  \citep[MESA,][]{Paxton2011} and stellar oscillations with the ADIPLS pulsation package \textbf{0.2} \citep{2008Ap&SS.316..113C}. 
Analysis in this manuscript was performed with \textbf{python 3.5.1} libraries scikit-learn \textbf{0.17.1} \citep{scikit-learn}, NumPy \textbf{1.11.0} \citep{van2011numpy}, matplotlib \textbf{1.5.1}  \citep{Hunter:2007}, biokit \textbf{0.3.2} \citep{biokit} and pandas \textbf{0.19.0} \citep{mckinney} as well as \textbf{R 3.3.2} \citep{R} and the R libraries magicaxis \textbf{2.0.0} \citep{magicaxis}, RColorBrewer \textbf{1.1-2} \citep{RColorBrewer}, parallelMap \textbf{1.3} \citep{parallelMap}, data.table \textbf{1.9.6} \citep{data.table}, ggplot2 \textbf{2.1.0} \citep{ggplot2}, GGally \textbf{1.2.0} \citep{GGally}, scales \textbf{0.4.0} \citep{scales} and  Corrplot \textbf{0.77}. 

\bibliographystyle{aasjournal.bst}
\bibliography{statistical_relations}

\appendix

\section{Seismic Definitions} 
\label{sec:sdefs}
We denote any frequency separation $S$ as the difference between a frequency $\nu$ of spherical degree $\ell$ and radial order $n$ and another frequency: 
\begin{equation} 
  S_{(\ell_1, \ell_2)}(n_1, n_2) \equiv \nu_{\ell_1}(n_1) - \nu_{\ell_2}(n_2).
\end{equation}

The large-frequency separation is defined as
\begin{equation} 
  \Delta\nu_\ell(n) \equiv S_{(\ell, \ell)}(n, n-1)
\end{equation}
and the small-frequency separation is
\begin{equation}
  \delta\nu_{(\ell, \ell+2)}(n) \equiv S_{(\ell, \ell+2)}(n, n-1).
\end{equation}

\citet{2003AA...411..215R} have demonstrated that taking the ratio of the \emph{local} large and small-frequency separations reduces the systematic offset introduced from improper modelling of the near-surface super-adiabatic region. This ratio is defined as: 

\begin{equation}    \label{eqn:LSratio}
  \mathrm{r}_{(\ell,\ell +2)}(n) \equiv \frac{\delta\nu_{(\ell, \ell+2)}(n)}{\Delta\nu_{(1-\ell)}(n+\ell)}.
\end{equation}
In addition, it was shown that the frequency-dependent offset can be somewhat mitigated by constructing ratios from five-point frequency separations and the \emph{local} large separation:
\begin{equation} 
  \mathrm{r}_{(\ell, 1-\ell)}(n) \equiv \frac{\mathrm{dd}_{(\ell,1-\ell)}(n)}{\Delta\nu_{(1-\ell)}(n+\ell)} \label{eqn:rnl}
\end{equation}
where the five point separations are defined as:
\begin{align} 
  \mathrm{dd}_{0,1} \equiv \frac{1}{8} \big[&\nu_0(n-1) - 4\nu_1(n-1) \notag\\
                                 &+6\nu_0(n) - 4\nu_1(n) + \nu_0(n+1)\big]\\ 
  \mathrm{dd}_{1,0} \equiv -\frac{1}{8} \big[&\nu_1(n-1) - 4\nu_0(n) \notag\\
                                &+6\nu_1(n) - 4\nu_0(n+1) + \nu_1(n+1)\big]. \label{eqn:dlast}
\end{align}

We calculate dozens of oscillation frequencies per star with the mode sets available dependent on the internal structure of an individual model. We thus determine a single representative value 
by following the prescription of \citet{2012AA...537A..30M}. In order to mimic how the oscillation spectra would appear in an observational data we weight weight all frequencies by their position in a Gaussian envelope with  full-width at half-maximum of $0.66\cdot\nu_{\max}{}^{0.88}$ and centered at the predicted frequency of maximum oscillation power $\nu_{\max}$.  We then calculate the weighted median of each variable, which we denote with angled parentheses (e.g.\ $\langle r_{1,0}\rangle$).

\section{Asteroseismic Scaling Relations}

\begin{equation}
\nu_{\rm max} \approx \frac{ M/M_{\sun}(T_{\rm eff}/T_{\rm eff,\sun})^{3.5}}{L/L_{\sun}} \nu_{\rm max,\sun} \: 
\label{equ:nmax}
\end{equation}

\begin{equation}
\Delta\nu \approx \frac{(M/M_{\sun})^{0.5}(T_{\rm eff}/T_{\rm eff,\sun})^{3}}{(L/L_{\sun})^{0.75}}
\Delta\nu_{\sun} \: 
\label{equ:dnu}
\end{equation}

\citet{2016MNRAS.460.4277G} have shown that a metallicity-dependent correction is required for the Equation \ref{equ:dnu} scaling relation.
The $\Delta\nu_{\sun}$ term can be replaced with a more appropriate reference value which can be calcuated according to:

\begin{equation}
\Delta\nu_{\rm{ref}}=\rm{A} \cdot e^{\lambda \rm{T_{eff}}/10^4K} \cdot (\cos(\omega \cdot \rm{T_{eff}}/10^4K+\phi))+B,	
\label{eq:corrfunc2}
\end{equation}
and where the unkown terms are listed in Table \ref{tab:pars2}.

\begin{table}
	\centering
	\caption{Parameters of the correction function.}
	\label{tab:pars2}
	\begin{tabular}{lc} 
		\hline
		A & 0.64$\cdot$[Fe/H] + 1.78  $\mu Hz$ \\
		$\lambda$ & $-$0.55$\cdot$[Fe/H] + 1.23  \\
		$\omega$ & 22.21 rad/K \\
		$\phi$ & 0.48$\cdot$[Fe/H] + 0.12 \\
		B & 0.66$\cdot$[Fe/H] + 134.92 $\mu Hz$ \\
		\hline
	\end{tabular}
\end{table}

\section{Correlation Plot} 
\label{sec:fullcorr}

\begin{figure*}
    \centering
    \includegraphics[trim={1cm 0 2cm 1cm},clip,
    width=0.95\textwidth]{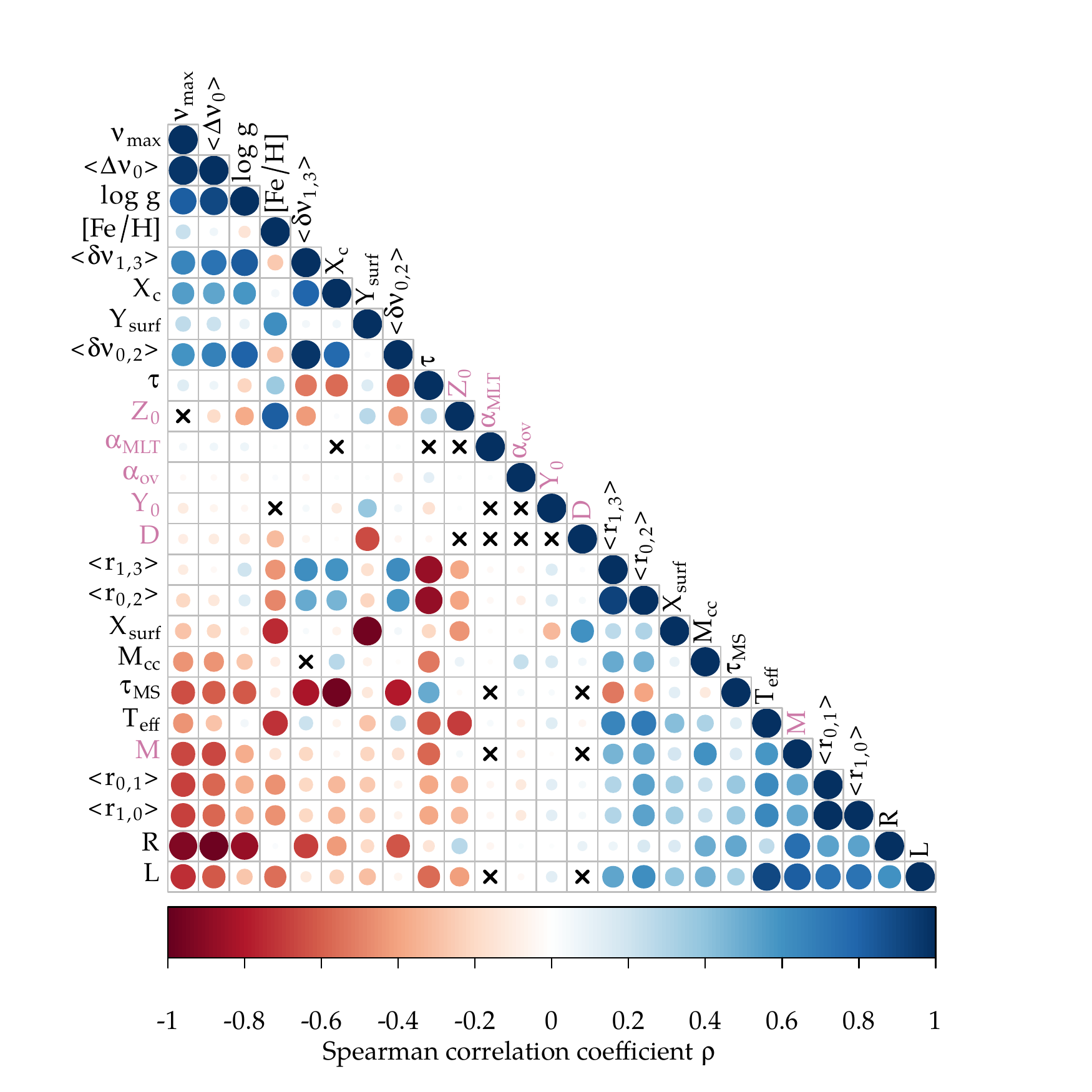}
    \caption{Spearman rank correlation matrix comprising various stellar and asteroseismic parameters. The quantities are as described in Table \ref{tab:parmdefs}
    with model input parameters marked in purple above. The complete grid of models are considered here. The size and color of each circle indicates the sign and magnitude of the Spearman rank coefficient, $\rho$, between two variables.  All correlations are significant excepting the entries indicated with a cross. The variables are ordered by the first principal component of the correlation matrix. }
    \label{fig:corr}
\end{figure*}

The full BA1 grid introduces some biases in our correlation analysis, particularly from tracks with calculated with  high-mass and/or high-diffusion. Correlation analysis with all models included are presented in Figure \ref{fig:corr}. 

A major difference that arises between Figure \ref{fig:filt_corr} and Figure \ref{fig:corr} is in the ordering of variables. Recall that we report the quantities according to the first principle component of the correlation matrix. Different combinations of variables are required to maximise the variance of each principle component in the new parameter space. Although the PCA analysis in Figures \ref{fig:corr-pcaobs} and Figures \ref{fig:corr-pcamods} rely on Pearson rather than Spearman correlations, they do demonstrate the difference in the composition of the PCs in each grid. 

We also find differences in the correlations that pertain to current surface abundance parameters. Consider the pair M -- Y$_{\rm{surf}}$. In Figure \ref{fig:corr} we find a small but non-negligible negative correlation. The reason being that higher mass tracks diffuse the helium from their surface more efficiently than low-mass stars.  Without the influence of these stars in our sample, our significance test yields a null correlation in Figure \ref{fig:filt_corr}; the expected result from a quasi-random distribution of initial abundances. 

Two interesting features emanating from our grid selection relates to the parameter pairs $\langle\delta\nu_{02}\rangle$ -- $T_{\rm{eff}}$ and $\langle r_{02}\rangle$ -- $\log \rm{g}$. 
We find a null correlation  between $\langle\delta\nu_{02}\rangle$ -- $T_{\rm{eff}}$ in truncated grid however this emerges as a small positive correlation when the full grid is considered. 
In \S \ref{sec:sages} we discussed the redundancy in the C-D diagram when projecting stellar models varied in six dimensions into a two-dimensional parameter space.  Thus the  null correlation 
arising from the truncated grid reflects the fact there many combinations of (primarily) mass and metallicity and hence temperature at a given age.  
The full grid, however, consists of a large number of hot short-lived stars that impart a noticeable trend. 

A similar argument applies to $\langle r_{02}\rangle$ -- $\log \rm{g}$. There are a great number of combinations of $\langle\Delta\nu_0\rangle$ and
$\langle\delta\nu_{02}\rangle$ for a given $\langle r_{02}\rangle$ thus in the truncated grid no correlation with $\log \rm{g}$ is registered. 
Once again the number of massive short-lived stars bias this previous null correlation.

Finally we note two minor results. Some pairs of parameters in the truncated grid which report null correlations in  Figure \ref{fig:filt_corr},  show very weak correlations in Figure \ref{fig:corr}. 
We refer to  L -- $\alpha_{\rm{ov}}$ and $\alpha_{\rm{MLT}}$ -- $\langle\delta\nu_{02}\rangle$ as cases in point. The correlations remain very weak in the current analysis and the larger sample size has introduced a minor trend that in this case passes our conservative significance criterion. We note also that most variables display a much stronger correlation with age in the full grid.

\section{Principle Component Analysis Explained Variance} 
\label{sec:fullPCA}

The PCs and their correlations will change depending on the number of dimensions included in the grid and the range of values each parameter takes; the PCs identify vectors of maximal variance. 
Our aim is to determine whether the PCs capture fundamental features ubiquitously encoded in the observables. 
Thus, we wish to investigate the information inherent to the dimensions and mitigate the impact of parameter ranges on our PCs. 
In order to provide a more robust interpretation we have calculated the PCs and their correlations with four different considerations given to the BA1 grid:
\begin{description}
    \item[\textbf{Grid A}] The full BA1 training grid;
    \item[\textbf{Grid B}] The truncated grid;
    \item[\textbf{Grid C}] A grid where more than half the models in each track possess metallicities of [Fe/H] $> -2$; and
     \item[\textbf{Grid D}] A grid with masses limited to $M < 1.2$ M$_{\sun}$.
\end{description}
Qualitative correlations between the stellar parameters and the PCs in each grid are presented in Figures \ref{fig:corr-pcaobs} and Figures \ref{fig:corr-pcamods}.

\begin{deluxetable*}{ccccccc}
\tablewidth{0pt}
\tablecaption{Percentage of the variance explained by each principle component. We report the explained variance percentages for the complete grid of training models (Grid A) and for the truncated set (Grid B, see \S \ref{sec:RCT}) that better encompasses the observational parameter space. In each case we consider the grid with and without the inclusion of $\nu_{\rm{max}}$ which is estimated using the \citet{1995AA...293...87K} scaling relations rather than calculated from first principle equations. We also consider the explained variances when limits are placed on the metallicity (Grid C) and mass (Grid D) ranges of the models.  These grids are used in \S \ref{sec:disc} to help interpret the PCs. \label{tab:PCAEV}} 
\tablehead{
        \colhead{}  &
        \multicolumn{4}{c}{$\nu_{\rm{max}}$ Included}               &
        \multicolumn{2}{c}{$\nu_{\rm{max}}$ Excluded}\\
        \colhead{Component}  &
        \colhead{Grid A} &
        \colhead{Grid B} &
        \colhead{Grid C} &
        \colhead{Grid D} &
        \colhead{Grid A} &
        \colhead{Grid B} 
}
\startdata
PC$_1$ 	&	41.79	&	42.36	& 42.49 & 42.74 &	40.89	&	41.47	\\
PC$_2$ 	&	36.12	&	34.18	& 37.49 & 35.89 &	36.52	&	33.65	\\
PC$_3$	&	9.17	&	11.65	& 9.39  & 10.25 &	8.99	&	12.21	\\
PC$_4$	&	7.69	&	9.79	& 7.69  &  6.89 &	8.27	&	10.58	\\
PC$_5$ 	&	4.23	&	1.23	& 2.14  &  3.36 &	4.55	&	1.36	\\
PC$_6$	&	0.54	&	0.48	&0.41    & 0.53   &	0.48	&	0.51	\\
PC$_7$	&	0.25	&	0.18	&0.24    &0.18    &	0.16	&	0.12	\\
PC$_8$ 	&	0.12	&	0.08	&0.09    &0.10    &	0.10	&	0.09	\\
PC$_9$	&	0.05	&	0.03	&0.04    &0.04    &	0.03	&	0.02	\\
PC$_{10}$ 	&	0.02	&	0.01 &0.01   &0.01	&	0.01	&	0.00	\\
PC$_{11}$	&	0.01	&	0.00 &0.00   &0.00     &	--	    &	--	
\enddata
 \end{deluxetable*} 

\section{PCA Correlation Analysis} \label{sec:ccoefs}
Figures \ref{fig:GCA-pcabar}a and \ref{fig:GCA-pcabar}b demonstrate the correlation strengths between our stellar parameters and the first five PCs. In Tables \ref{tab:ocoefs} and \ref{tab:ocoefs} we list the coefficients between all parameters and all PCs. The table is useful for  determining whether the transitive criterion applies to parameters within a given PC. It also aids in the calculation of the $\Lambda$ scores in \S \ref{sec:ISP}.

\begin{table*}[]
\centering
\caption{Pearson's $r$ coefficients between the principle components and observables in the truncated grid.}
\label{tab:ocoefs}
\begin{tabular}{c|ccccccccccc}
\hline \hline
& $\log \rm{g}$ &$T_{\rm{eff}}$  & $[\rm{Fe/H}]$ & $\langle\Delta\nu_0\rangle$  & $\langle\delta\nu_{02}\rangle$ & $\langle r_{02}\rangle$    & $\langle r_{01}\rangle$    & $\langle\delta\nu_{13}\rangle$ &  $\langle r_{13}\rangle$   &  $\langle r_{10}\rangle$   & $\nu_{\rm{max}}$        \\ \hline
PC$_1$    & 0.93  & -0.20 & -0.35 & 0.92  & 0.93  & 0.38  & -0.07 & 0.95  & 0.32  & -0.07   & 0.87  \\
PC$_2$    & -0.30 & 0.73  & -0.29 & -0.33 & 0.34  & 0.85  & 0.81  & 0.22  & 0.76  & 0.81    & -0.42 \\
PC$_3$    & 0.00  & -0.60 & 0.63  & 0.04  & 0.06  & 0.15  & 0.45  & -0.13 & -0.17 & 0.45    & 0.22  \\
PC$_4$    & 0.08  & 0.08  & -0.60 & 0.19  & -0.08 & -0.30 & 0.37  & -0.13 & -0.52 & 0.37    & 0.10  \\
PC$_5$    & 0.14  & 0.25  & 0.20  & 0.06  & -0.02 & -0.05 & 0.03  & 0.00  & -0.07 & 0.03    & 0.01  \\
PC$_6$    & 0.11  & -0.01 & -0.04 & 0.01  & -0.11 & 0.09  & 0.00  & -0.12 & 0.06  & 0.00    & 0.04  \\
PC$_7$    & -0.09 & 0.03  & 0.01  & 0.05  & -0.03 & -0.01 & 0.00  & -0.01 & 0.03  & 0.00    & 0.09  \\
PC$_8$    & 0.02  & -0.02 & 0.00  & 0.00  & -0.04 & -0.05 & 0.02  & 0.03  & 0.05  & 0.02    & -0.01 \\
PC$_9$    & 0.01  & 0.01  & 0.00  & -0.04 & 0.02  & -0.02 & 0.00  & -0.02 & 0.01  & 0.00    & 0.03  \\
PC$_{10}$   & 0.00  & 0.00  & 0.00  & 0.02  & 0.02  & -0.01 & 0.00  & -0.02 & 0.01  & 0.00    & -0.01 \\
PC$_{11}$   & 0.00  & 0.00  & 0.00  & 0.00  & 0.00  & 0.00  & 0.01  & 0.00  & 0.00  & -0.01   & 0.00  \\ \hline
\end{tabular}
\end{table*}

\begin{table*}[]
\centering
\caption{Pearson's $r$ coefficients between the principle components and model parameters in the truncated grid.}
\label{tab:mcoefs}
\begin{tabular}{c|cccccccccccccc}
\hline \hline
    &$M$   & $Y$         & $Z$         & $\alpha_{\rm{MLT}}$     & $\alpha_{\rm{ov}}$ & $D$ & $\tau$      & $\tau_{\rm{MS}}$     & $X_c$      & $M_{\rm{cc}}$  & $X_{\rm{surf}}$   & $Y_{\rm{surf}}$   & $R$    & $L$                  \\ \hline
PC$_1$  & -0.67 & -0.08 & -0.30 & 0.06      & -0.05     & -0.12 & -0.19 & -0.72 & 0.77     & -0.45   & -0.04   & 0.16   & -0.86 & -0.69 \\
PC$_2$  & 0.29  & 0.10  & -0.41 & -0.15     & -0.04     & -0.17 & -0.56 & -0.26 & 0.14     & 0.17    & 0.05    & 0.11   & 0.30  & 0.55  \\
PC$_3$  & 0.13  & 0.04  & 0.48  & -0.10     & -0.07     & -0.03 & -0.01 & -0.04 & 0.14     & -0.11   & -0.33   & 0.17   & 0.09  & -0.04 \\
PC$_4$  & -0.51 & -0.17 & -0.40 & 0.03      & -0.09     & -0.08 & 0.51  & 0.48  & -0.49    & -0.34   & 0.29    & -0.17  & -0.23 & -0.13 \\
PC$_5$  & -0.02 & 0.17  & -0.09 & 0.44      & -0.16     & -0.43 & 0.13  & 0.13  & -0.17    & -0.29   & -0.50   & 0.59   & -0.14 & -0.03 \\
PC$_6$  & -0.02 & -0.20 & 0.08  & -0.01     & -0.15     & -0.10 & 0.28  & 0.07  & -0.02    & -0.25   & 0.05    & -0.10  & -0.14 & -0.03 \\
PC$_7$  & 0.09  & 0.05  & -0.05 & 0.14      & 0.11      & -0.15 & -0.10 & 0.03  & 0.10     & 0.30    & -0.14   & 0.17   & 0.18  & 0.38  \\
PC$_8$  & -0.11 & 0.08  & -0.06 & -0.25     & 0.05      & -0.07 & 0.03  & -0.03 & -0.02    & -0.12   & -0.05   & 0.08   & -0.07 & -0.09 \\
PC$_9$  & 0.21  & -0.35 & 0.22  & 0.20      & -0.06     & -0.08 & -0.22 & -0.12 & 0.15     & 0.04    & -0.01   & -0.10  & 0.06  & 0.07  \\
PC$_{10}$ & -0.17 & 0.26  & -0.14 & -0.15     & 0.02      & 0.06  & 0.17  & 0.12  & -0.10    & -0.01   & 0.01    & 0.06   & -0.09 & -0.04 \\
PC$_{11}$ & -0.01 & -0.01 & -0.01 & 0.01      & 0.01      & 0.00  & 0.00  & 0.00  & 0.00     & -0.02   & 0.01    & -0.01  & 0.00  & -0.01 \\ \hline
\end{tabular} 
\end{table*}

\section{PC correlations with different grids}
\label{sec:PCAg}
\begin{figure*}
    \centering
    \includegraphics[angle=90,width=0.7\textwidth]{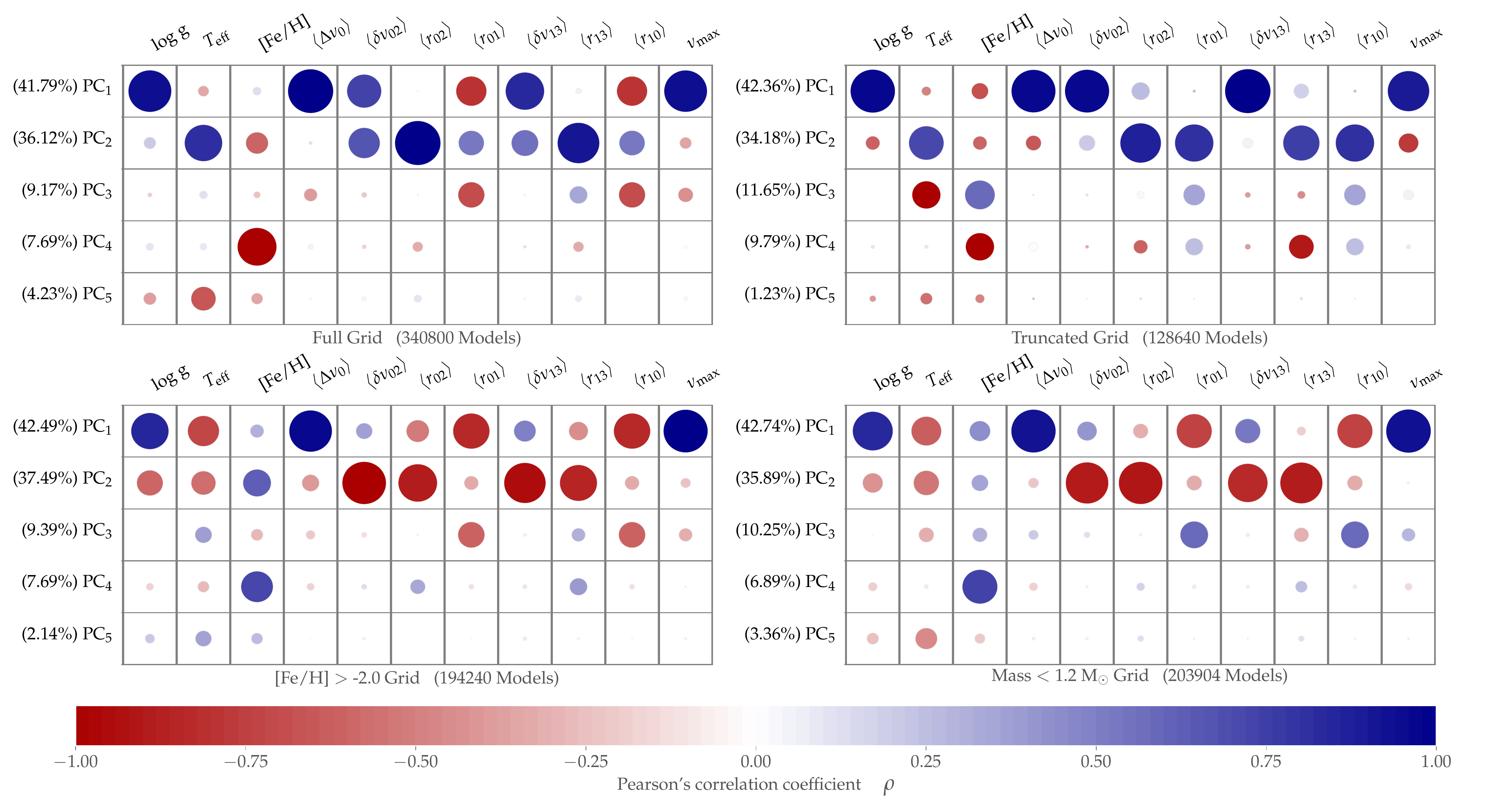}
    \caption{Pearson correlation matricies relating the principle components back to the stellar observables in each of the four grids described in Appendix \ref{sec:fullPCA}. } 
    \label{fig:corr-pcaobs}
\end{figure*}

\begin{figure*}
    \centering
    \includegraphics[angle=90,width=0.7\textwidth]{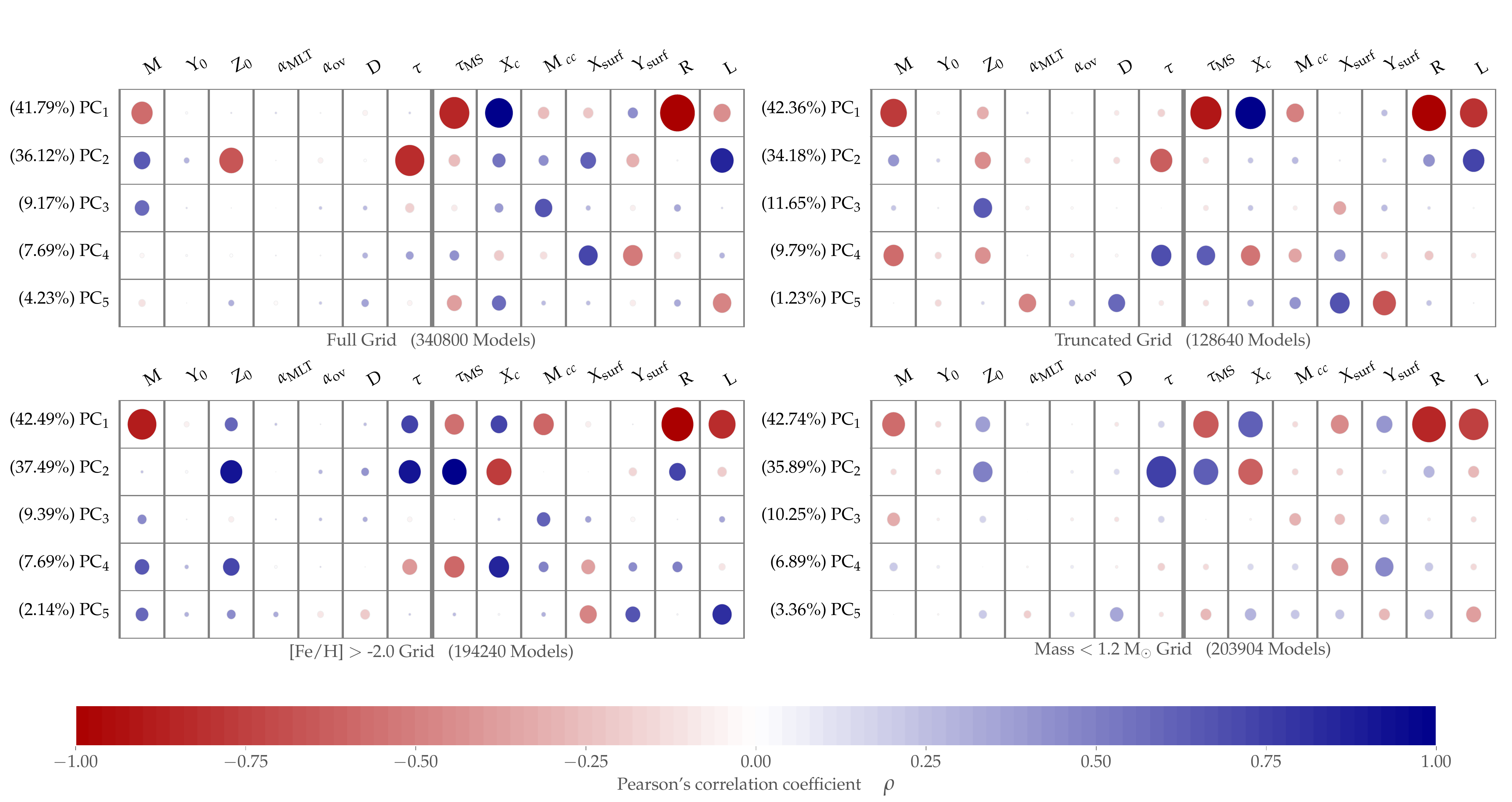}
    \caption{Pearson correlation matricies relating the principle components back to the model quantities in each of the four grids described in Appendix \ref{sec:fullPCA}.} 
    \label{fig:corr-pcamods}
\end{figure*}

In \S \ref{sec:intPC}  we presented the correlation strengths between the PCs and observables (Figure \ref{fig:GCA-pcabar}a) and the PCs and the model parameters (\ref{fig:GCA-pcabar}b). 
 Here we perform the same analysis with the different subsets of the BA1 grid described in Appendix \ref{sec:fullPCA}. In order to compare the results for each grid, in Figures \ref{fig:corr-pcaobs} and \ref{fig:corr-pcamods} we employ a correlation plot rather than the quantitative bar chart used in \S \ref{sec:intPC}. This allows an inspection of the qualitative behaviour of the PCs in each case. We find a similar explained variance from the corresponding PCs in each grid. This suggests that the PCs capture essentially the same inherent features in model data and that the PCs are not due to the number of models in our analysis or the chosen parameter ranges.

\section{$\Lambda$ Analysis}
\label{sec:lambdaa}

The data matrix of observables $\mathbf{X}$ is size $n \times p$ where n is the number of training models and p the number of parameters.
We centre and scale the entries according to the mean and standard deviation of each parameter.
The resultant matrix,  $\mathbf{\bar{X}}$, therefore has the property that for each parameter, $p$, $\mu(p) =0$ and $\sigma(p) =1$. 
We compute the correlation matrix, $R$, for the matrix  $\mathbf{\bar{X}}$ :
\begin{eqnarray}
    \mathbf{R}  = &\rm{Corr}(\mathbf{\bar{X}}) \\[3pt] \nonumber
                = & \mathbf{\bar{X}}\mathbf{\bar{X}^\top}.  
\end{eqnarray}
As the correlation and covariance matrices are symmetric we calculate the eigendecomposition of R such that:
\begin{equation}
\mathbf{R}=\mathbf{VLV^\top},
\end{equation}  
where \textbf{V} a matrix of eigenvector columns and \textbf{L} a diagonal matrix of eigenvalues.
The eigenvectors specify the principal axes of the data and the eigenvalues indicate the amount of variance there is in the data in the direction of the corresponding eigenvector.
We can define the projection matrix \textbf{P} such that we project/transform our data into the new space    
\begin{equation}
\mathbf{P} = \mathbf{\bar{X}}  \mathbf{V}.
\end{equation}

The correlation matrix is a special case of the covariance matrix in that the former is normalised.
For generality let us consider the covariance matrix, such that the original data matrix was centred but not scaled ($\mathbf{\hat{X}}$), then
\begin{eqnarray}
    \mathbf{C}  = &\rm{Cov}(\mathbf{\hat{X}}) \\[3pt] \nonumber
                = & \frac{1}{n-1}\mathbf{\hat{X}}\mathbf{\hat{X}^\top}\\[3pt] \nonumber
                = & \mathbf{V} \mathbf{L} \mathbf{V^\top},
\end{eqnarray}
where we divide by (n-1) to unbias to covariance (the covariance entries will have different scales).

Alternatively and equivalently, we may extract our PCs through SVD of  $\mathbf{\hat{X}}$ such that: 
\begin{equation}
    \mathbf{\hat{X}} = \mathbf{U}  \mathbf{\Sigma} \mathbf{ V^\top}
\end{equation}
where \textbf{U} is the left matrix of singular orthogonal vectors with dimensions $n \times n$,
$\mathbf{\Sigma}$ is a diagonal matrix of singular values with dimensions $n \times p$, 
and $\mathbf{V^{\top}}$ is the right matrix of singular  orthogonal vectors with diemsnions $p \times p$. 
The diagonal elements of $\mathbf{\Sigma}$ assign a  relative  importance  to  each  vector whereas the vectors of \textbf{V} are the principal directions/axes.
As the matricies \textbf{U} and \textbf{V} comprise orthogonal components they have the property
\begin{eqnarray}
\label{eqn:ident}
\mathbf{U^{\top}U}=\mathbf I_{n \times n} \\ \nonumber
\mathbf{V^{\top}V}=\mathbf I_{p \times p}.
\end{eqnarray}
We note also that 
\begin{eqnarray}
&\left(\mathbf {A \cdot B \cdot C }\right)^\top = \mathbf{C^\top \cdot B^\top \cdot A^\top} \\
&\implies  (\mathbf U\mathbf \Sigma\mathbf V^\top)^\top = (\mathbf V\mathbf \Sigma\mathbf U^\top)
\end{eqnarray}
as $\mathbf{\Sigma}$ is a diagonal matrix.

We can reconstruct the eigendecomposition of the covariance matrix from the SVD: 
\begin{eqnarray}
 \frac{1}{n-1}\mathbf{\hat{X}}\mathbf{\hat{X}}^\top = & \frac{1}{n-1} (\mathbf U\mathbf \Sigma\mathbf V^\top)(\mathbf U\mathbf \Sigma\mathbf V^\top)^\top\\[3pt] \nonumber
=& \frac{1}{n-1}(\mathbf U\mathbf \Sigma\mathbf V^\top)(\mathbf V\mathbf \Sigma\mathbf U^\top)
\end{eqnarray}
and from our identities in Equation \ref{eqn:ident}
\begin{equation}
 \frac{1}{n-1}\mathbf{\hat{X}}\mathbf{\hat{X}}^\top=\mathbf U \frac{\mathbf \Sigma^2}{n-1} \mathbf U^\top.
\end{equation}
We therefore find that the square roots of the eigenvalues of $\mathbf{C}$ are the singular values of  $\mathbf{\bar{X}}$ and that the vectors in the right singular matrix, \textbf{V}, are the principal directions/axes. The projection matrix can be calculated from the SVD such that 
\begin{eqnarray}
\mathbf{P} = &\mathbf{\hat{X}}  \mathbf{V} \\[3pt] \nonumber
           = & \mathbf U \mathbf \Sigma \mathbf V^\top \mathbf{V}  \\[3pt] \nonumber
           = & \mathbf U \mathbf \Sigma.
\end{eqnarray}

The PCA loadings are the columns of \textbf{L} which implies that 
\begin{equation}
\mathbf{L}=\mathbf{V}\frac{\mathbf \Sigma}{\sqrt{n-1}}.
\end{equation}
We can see that the loadings are the eigenvectors scaled by the square roots of the respective eigenvalues.
With these definitions we can compute the cross-covariance matrix between original variables and the standardized projection matrix. 
To calculate the standardized PC scores for \textbf{P} we require each column of \textbf{U} to have unit variance. As $\mathbf{\Sigma}$ is diagonal it is simply a scaling matrix and can be dropped here yielding: 
\begin{eqnarray}
&\frac{1}{n-1}\mathbf{X}^\top(\sqrt{n-1}\mathbf{U}) = \\ &\frac{1}{\sqrt{n-1}}\mathbf{V}\mathbf{\Sigma}\mathbf{U}^\top\mathbf{U} =\\ &\frac{1}{\sqrt{n-1}}\mathbf{V}\mathbf{\Sigma}=\mathbf{L}.
\end{eqnarray}
We find that the covariance matrix between the standardized PCs and original variables is in fact given by the loadings. 
In  \S \ref{sec:ev} we computed the \emph{correlations} between the observables and their PCs rather than the covariances, requiring that the observables are normalized by their standard deviation. As we centred and scaled our data prior to performing the PCA, their values are unity and our correlation analysis is therefore equivalent to reporting the loadings. 

The correlation analysis allowed us to project the model data onto the PC space and determine the `equivalent' loadings for each parameter. Through the $\lambda$ score we can therefore determine to what extent the variance in the model data is captured by the PCs.  In Table \ref{tab:corrLFull} we compare the results of the analysis for each grid. We find similar results for most parameters 
with differences in some of the initial model parameters due to their underlying distributions as a result of the grid truncations.

\begin{table}[h]
    \centering
    \begin{tabular}{ccccc}
    \hline \hline
    & \multicolumn{4}{c}{$\Lambda_{\rm{param}}$} \\
Parameter & Grid A & Grid B & Grid C & Grid D \\ \hline     
R	&	0.97	&	0.97	&	0.98	&	0.97	\\
L	&	0.93	&	0.96	&	0.93	&	0.95	\\
$X_c$	&	0.93	&	0.94	&	0.93	&	0.94	\\
$\tau_{\rm{MS}}$	&	0.93	&	0.93	&	0.93	&	0.94	\\
M	&	0.91	&	0.91	&	0.92	&	0.88	\\
$\tau$	&	0.74	&	0.79	&	0.78	&	0.76	\\
Z$_0$	&	0.76	&	0.73	&	0.78	&	0.80	\\
M$_{cc}$	&	0.58	&	0.61	&	0.68	&	0.41	\\
Y$_{\rm{surf}}$	&	0.48	&	0.50	&	0.55	&	0.54	\\
X$_{\rm{surf}}$	&	0.50	&	0.48	&	0.53	&	0.55	\\
$\alpha_{\rm{MLT}}$	&	0.02	&	0.38	&	0.04	&	0.06	\\
Y$_0$	&	0.10	&	0.31	&	0.27	&	0.09	\\
D	&	0.13	&	0.29	&	0.22	&	0.21	\\
$\alpha_{\rm{ov}}$	&	0.10	&	0.08	&	0.11	&	0.12	\\

 \hline
    \end{tabular}
    \caption{The $\Lambda$ score is a sum of the squares of  $r(X, PC_i)$ indicating the variance explained for a given parameter. These scores are by definition unity for our observables.}
    \label{tab:corrLFull}
\end{table}

\section{The Impact of Uncertainties for Upcoming Photometric Space Missions} 
\begin{figure*}
    \centering
    \includegraphics[width=0.95\textheight,height=0.95\textwidth,keepaspectratio, angle=90]{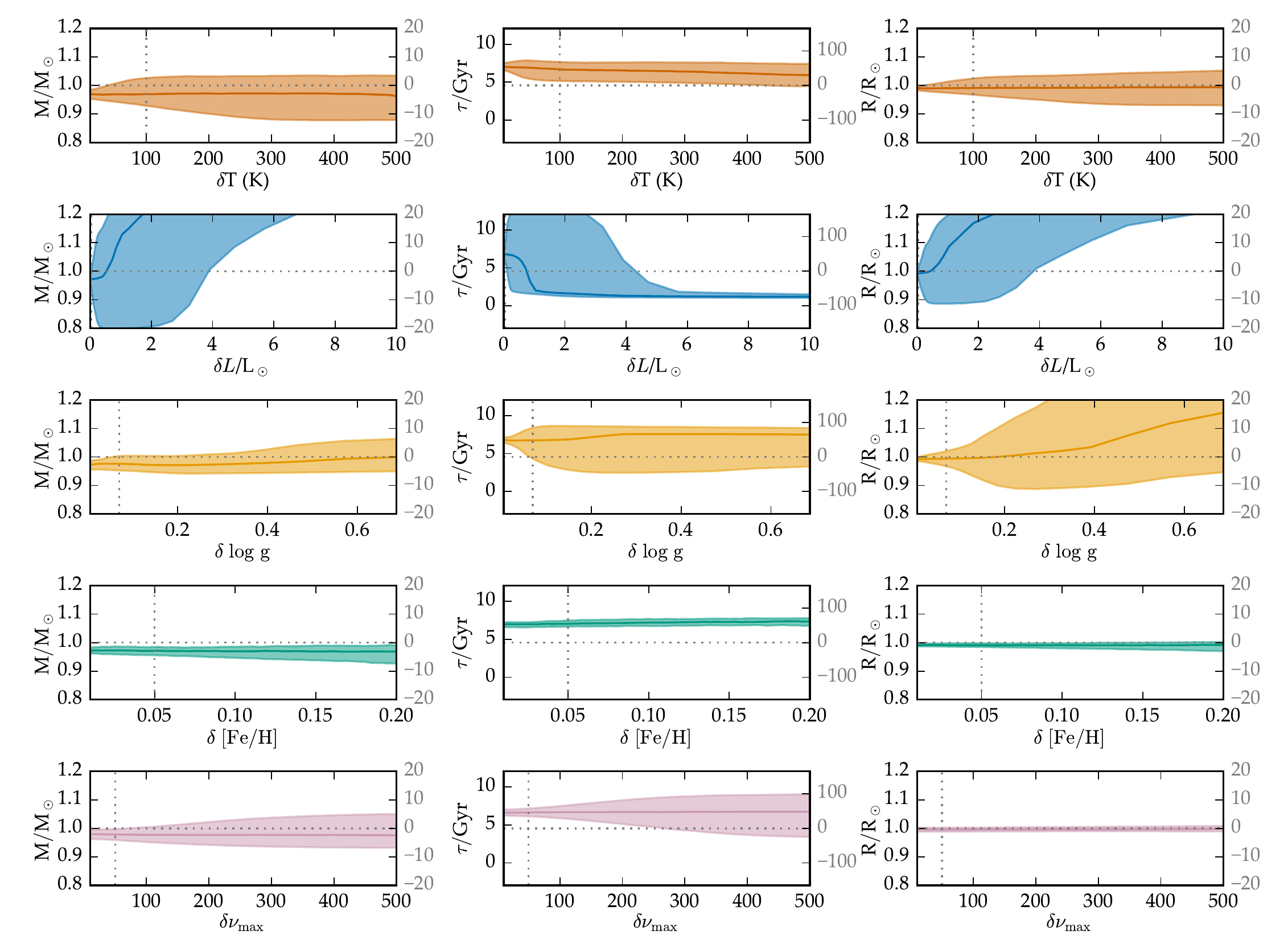}
    \caption{Predictions for the solar mass, age, luminosity and radius as a function of the uncertainties applied to key observables. In each panel we have perturbed the quantity on the abscissa in isolation, centred around the measured value listed in  Table \ref{tab:sunstar} and with the uncertainties in the ranges specified therein. We indicate the median predicted value (solid line) and the 68\%  confidence interval (shaded region). Here the observables comprise those expected from the \emph{TESS} space mission assuming that the p-mode power excess can be extracted.} 
    \label{fig:uncerttess}
\end{figure*}

\begin{figure*}
    \centering
    \includegraphics[width=0.95\textheight,height=0.95\textwidth,keepaspectratio, angle=90]{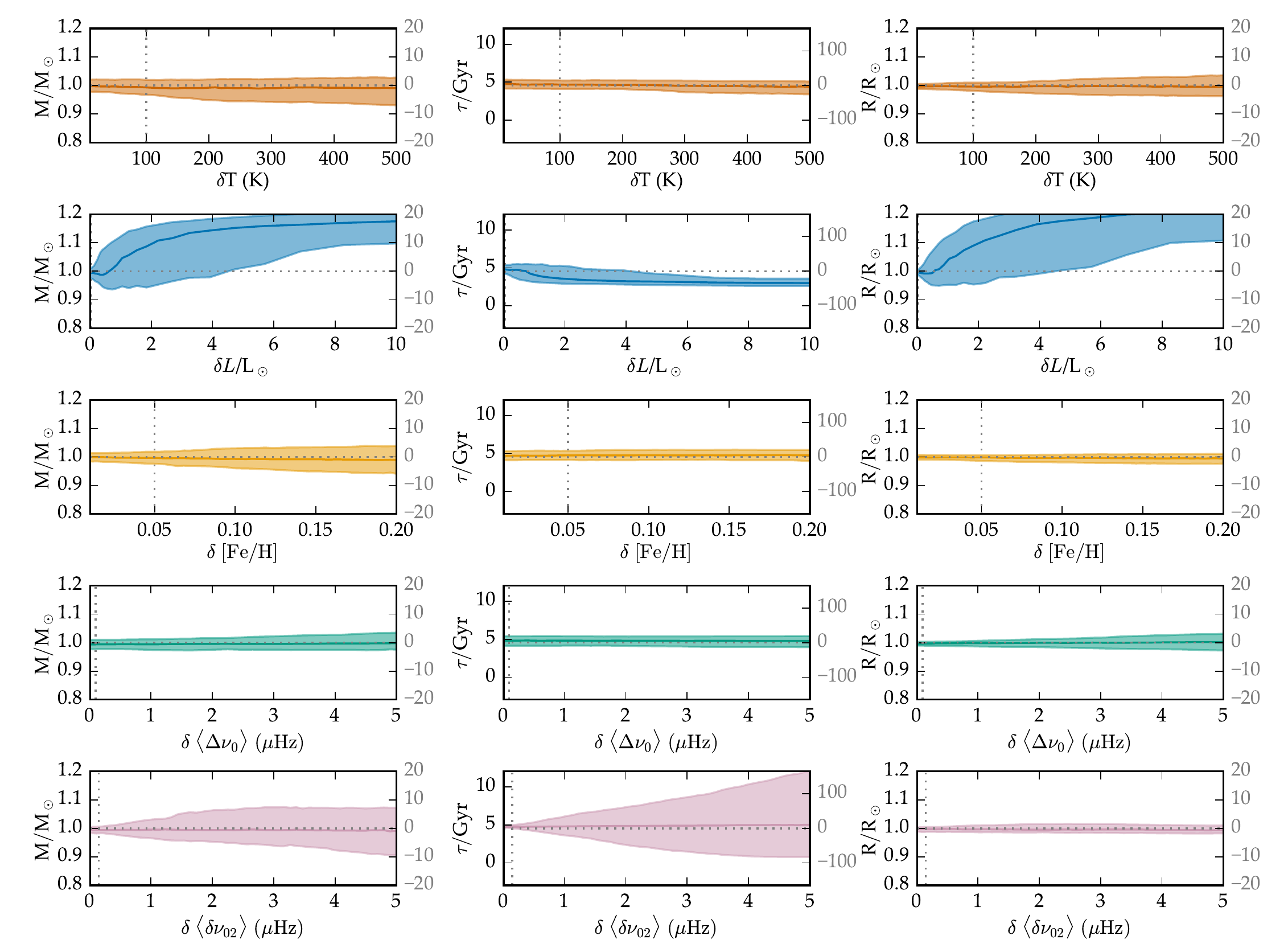}
    \caption{Predictions for the solar mass, age, luminosity and radius as a function of the uncertainties applied to key observables. In each panel we have perturbed the quantity on the abscissa in isolation, centred around the measured value listed in  Table \ref{tab:sunstar} and with the uncertainties in the ranges specified therein. We indicate the median predicted value (solid line) and the 68\%  confidence interval (shaded region). Here the observables comprise those expected from the \emph{PLATO} space mission.} 
    \label{fig:uncertplato}
\end{figure*}

Below we demonstrate the impact of measurement uncertainty on the prediction of parameters from the upcoming \emph{TESS} (Figure \ref{fig:uncerttess}) and \emph{PLATO} (Figure \ref{fig:uncertplato}) space missions.  We produce  probability density distributions for 250 sets of $\sigma$ values for each parameter we predict. The ranges for each parameter from which we draw our $\sigma$ values are listed in Table \ref{tab:sunstar}. We restrict out observables to those we are likely to possess from the respective missions. 
In each figure we plot the median value (solid line) and the 68\%  confidence interval (shaded region).
 \begin{table}
    \caption{Central solar values and uncertainty ranges used for predictions in Figures \ref{fig:uncerttess} and \ref{fig:uncertplato}.}
    \begin{tabular}{lcccccc}
    \hline \hline
 \multicolumn{1}{c}{} & \multicolumn{3}{c}{TESS}& \multicolumn{3}{c}{PLATO}  \\
Quantity & Value & Min($\sigma$) & Max($\sigma$) & Value & Min($\sigma$) & Max($\sigma$)\\ \hline 
$T_{\rm{eff}}$ (K)  & 5777 & 10  & 500& 5777 & 10  & 500\\
$\log \rm{g}$ &  4.44 & 0.0001 & 1.0&  4.44 & 0.0001 & 1.0\\
$[\rm{Fe/H}]$ & 0.0 & 0.05 & 0.5 & 0.0 & 0.05 & 0.5\\
$L$ & 1.0 & 0.001 & 10 & 1.0 & 0.001 & 10 \\
$\nu_{\rm{max}}$ & 3050 & 10 & 500 & -- & -- & --\\
$\langle\Delta\nu_0\rangle$ ($\mu$Hz) & -- & -- & --& 136.0 & 0.5 & 50\\
$\langle\delta\nu_{02}\rangle$ ($\mu$Hz)& -- & -- & -- & 9.0 & 0.5 & 5 \\
\hline
    \end{tabular}
    \label{tab:sunstar}
\end{table}

\label{sec:PSM}

\end{document}